\documentclass[twocolumn, pre]{revtex4-1}
\usepackage[intlimits]{amsmath}
\usepackage{amssymb}
\usepackage{graphicx}
\usepackage{epstopdf}
\usepackage{dcolumn}
\usepackage{hyperref}

\usepackage{natbib}
\def\posVec{{\bf r}}
\def\PosVec{{\bf R}}
\def\bForce{{\bf q}}

\begin{document}
\title{\bf Solute drag forces from equilibrium interface fluctuations}
\author{Changjian Wang and Moneesh Upmanyu}
\email{mupmanyu@neu.edu}
\affiliation
{
Group for Simulation and Theory of Atomic-Scale Material Phenomena ({\it st}AMP)\\
Department of Mechanical and Industrial Engineering, Northeastern University, Boston, MA 02115. 
}
\begin{abstract}
The design of polycrystalline alloys hinges on a predictive understanding of the interaction between the diffusing solutes and the motion of the constituent crystalline interfaces. Existing frameworks ignore the dynamic multiplicity of and transitions between the interfacial structures and phases. Here, we develop a computationally-accessible theoretical framework based on short-time equilibrium fluctuations to extract the drag force exerted by the segregating solute cloud. Using three distinct classes of computational techniques, we show that the random walk of a solute-loaded interface is necessarily non-classical at short time-scales as it occurs within a confining solute cloud. The much slower stochastic evolution of the cloud allows us to approximate the short-time behavior as an exponentially sub-diffusive Brownian motion in an external trapping potential with a stiffness set by the average drag force. At longer time-scales, the interfacial and bulk forces lead to a gradual recovery of classical random walk of the interface with a diffusivity set by the extrinsic mobility. The short-time response is accessible via {\it ab-initio} computations, offering a firm foundation for high throughput, rational design of alloys for controlling microstructural evolution in polycrystals, and in particular for nanocrystalline alloys-by-design.   
\end{abstract}
\maketitle
\section{Introduction}
Engineering the chemistry and concentration of solutes in crystalline materials is an economical route for tailoring their structural and functional properties. The solutes modify the free energy of bulk phases, and appropriately designed processing routes can readily exploit their relative stability for the design of multiphase microstructures~\cite{gg:Smith:1948}. A more dramatic effect involves their interactions with extended defects such as dislocations and crystalline interfaces. The defect microstructure serves as a natural template for the inhomogeneous distribution of solutes as they preferentially segregate at these less dense regions~\cite{disl:Cottrell:1948, gbm:Gottstein:1998, book:SuttonBalluffi:1995}. Their kinetics is modified as it is coupled to the diffusion of the enveloping solute cloud.
%and this has far reaching consequences. 
As a well-known example, the Cottrell (solute) cloud around dislocations in metals arrests their motion~\cite{disl:CottrellBilby:1950}. The effect is the basis for solid-solution strengthening and strain aging in a range of industrially relevant alloys~\cite{disl:Baird:1971}. 

Solute segregation at interfaces is equally important during annealing phenomena such as grain growth and recrystallization, and more general thermo-mechanical processing~\cite{gbseg:Hondros:1965, gbseg:Balluffi:1979}. The segregation modifies the interfacial thermodynamics and kinetics, and thereby the mechanical and functional properties of interfacial microstructures. As examples, segregation-based changes in the interfacial free energy can lead to transitions in its structure and morphology~\cite{intfacetimpurity:DonaldBrown:1979, gbt:Cahn:1982,gbt:Shvindlerman:1985,intfacetimpurity:FerenceBalluffi:1989,gbseg:Kirchheim:2002, gbt:Frolov:2013}, in turn modifying the driving force for coarsening~\cite{gbseg:Weissmuller:1993, gbseg:Schuh:2012}. Drag forces exerted by the usually sluggish solute cloud~\cite{imdrag:LuckeDetert:1957,imdrag:Cahn:1962, imdrag:Machlin:1962,imdrag:HillertSundman:1976,imdrag:GlaeserCannon:1986,imdrag:MendelevSrolovitz:2002,imdrag:MaDregiaWang:2003,imdrag:KorzhenevskiiBauschSchmitz:2006,imdrag:HersentNes:2006} lead to a dramatic decrease in the interface kinetics~\cite{gbm:Gottstein:1998,gbm:TrauttUpmanyu:2006}. The coupled evolution of the interfacial microstructure leads to changes in the final grain size, texture and distribution of interface types that directly impact their material properties~\cite{imdrag:Michels:1999,imdrag:GlaeserCannon:1986b,imdrag:BerryHarmer:1986}. 
The ability to quantify the segregation extent and the drag forces is therefore of central importance during processing of polycrystalline alloys and related multiphase materials. 

 \subsection{Theoretical Background}
Extracting the segregation extent and its effect on the interface kinetics is challenging as the parameters are sensitive to the dynamic atomic-scale structure of the interfaces. 
%To see this, it is instructive to revisit the salient aspects of existing theoretical frameworks for solute-drag~\cite{imdrag:LuckeDetert:1957,imdrag:Cahn:1962,gbm:LuckeStuwe:1971,imdrag:Machlin:1962,imdrag:HillertSundman:1976}.
The segregation-driven coupling between interfaces and the solute cloud is controlled by the strength of the interaction between each solute and the interface~\cite{imdrag:LuckeDetert:1957,imdrag:Cahn:1962,gbm:LuckeStuwe:1971,imdrag:Machlin:1962,imdrag:HillertSundman:1976} $U(\PosVec)$, where $\PosVec$ is the generalized spatial coordinate. In general, this strength varies differently along and normal to the interface, $U\equiv U(\posVec, z)$ where $\posVec\equiv(x,y)$ is the coordinate within the interface plane and $z$ the coordinate along the interface normal. For unsaturated interfaces in dilute ideal solutions, this thermodynamic parameter sets the equilibrium segregation isotherm for the stationary interface, 
\begin{align}
\label{eq:equibConc}
c(\PosVec)\approx c_\infty e^{-U(\posVec, z)/k_BT}\,, 
\end{align} 
where $c_\infty$ is the bulk impurity concentration far away from the interface.
%and $\PosVec \equiv (\posVec, z)$ is the generalized spatial coordinate. 

The solute-drag effect arises due to the modified chemical potential of the solutes in the vicinity of the interface. Following early work by Cahn, L\"{u}cke and St\"{u}we on grain boundaries in ideal dilute solutions (CLS framework)~\cite{imdrag:LuckeDetert:1957,imdrag:Cahn:1962}, 
\begin{align}
\mu(\posVec, z) - \mu_0= k_BT\ln c(\PosVec, t) + U(\posVec, z), \nonumber% + \Omega\gamma\kappa.
\end{align}
with $\mu_0$ the reference chemical potential corresponding to bulk composition $c=c_\infty$.  
%although regular solution corrections can be readily incorporated. 
The local diffusive flux is ${\bf J} =({D_s c}/{k_BT})\, \nabla \mu$ is controlled by the solute diffusivity $D_s \equiv D_s(z)$ that in general varies with the distance away from the boundary. Usually, the boundary diffusivity $D_s^{b}=D(z<|\lambda|)$ over the nanoscopic boundary width $\lambda$  is much faster than the bulk diffusivity $D_s^B=D(z>|\lambda|)$ such that the flux is largely determined by fast diffusion across the boundary region. In this limit, the temporal evolution of the profile takes the form
\begin{align}
\frac{\partial c}{\partial t} = D_s^b \Delta c  + \frac{D_s^b}{k_BT} \nabla \cdot (c \nabla U)\nonumber.
\end{align}

The in-plane variations in $U(\posVec, z)$ reflect the structural periodicity of the interface that lead to short-range fluctuations in individual solute-interface interactions. 
%To make analytical progress, the CLS framework
Averaging these out such that $U\equiv U(z)$ simplifies solutions for the shape of a steady-state solute cloud around a flat interface moving at a constant velocity $\bar{h}_t$ along its normal $\hat{\bf n}$,  
%For this simplified scenario, 
%The shape of the cloud follows from a 1D flux balance in a coordinate system moving with the boundary,  
\begin{align}
\hat{\bf n} \cdot {{\bf J}} = D_s^b \frac{\partial c}{\partial z}  + \frac{D_s^b c}{k_BT}\frac{\partial U}{\partial z} + \bar{h}_t c = \bar{h}_t c_\infty.\nonumber
\end{align}
%\left[1- c \right]
As expected, the solution for the shape of the solute cloud $c(z)-c_\infty$ depends on both the interface velocity and the variation $U(z)$. The net drag force then is the  
%The last term is the Gibbs thompson effect that modifies the
%The spatial gradient of the interaction strength is 
force (pressure) per solute on the interface 
%$p_{s/s} (\posVec, z) = -\nabla U(\posVec, z)$
$\nabla U(\posVec, z)$, integrated over all the solutes within the cloud,
\begin{align}
\label{eq:mcForce}
%p_i=N \int_{\partial\mathcal{P}} c \; \nabla U\,dz,
p_s (t) &= \frac{1}{\lambda}\int_{-\lambda/2}^{+\lambda/2} \left[ \frac{1}{A}\int_{\Omega_A} \Gamma(\PosVec,t)  \nabla U(\posVec, z)\,d\posVec\right] dz\,,
%&= \int_{-\lambda/2}^{+\lambda/2} N_s\,[c(\PosVec, t)-c_\infty]  \nabla U(\posVec, z)\,d\PosVec
\end{align}
where $A$ is the interface area and $\lambda$ is the boundary width associated with the solute excess per unit interface area, defined as $d\Gamma =  [c(\PosVec, t)-c_\infty]\, dz$. 
The drag force renormalizes the relation between the interface velocity and the applied driving force. For a linear relation between the two, we get an extrinsic mobility that depends on the bulk solute concentration and the interface velocity, $M_i^\ast\equiv M_i^\ast(c_\infty, \bar{h}_t)$. Below a threshold velocity and for a symmetrical triangular profile for $U(z)$ centered around a flat boundary~\cite{imdrag:Cahn:1962}, 
\begin{align}
\label{eq:ExMobCLS}
\frac{1}{M_i^\ast} \approx \frac{1}{M_i} + \alpha c_\infty,
\end{align} 
where the parameter $\alpha$ is an integral related to the spatial variation in $U(z)$ and varies inversely with the solute diffusivity $D_s$.
%\begin{align}
%\label{eq:alphaCLS}
%\alpha &= 4 N_v k_BT \int_{-\infty}^\infty \frac{1}{D(z)} \sinh^2\left(\frac{U}{k_BT} \right)\,dz,\nonumber\\
%\alpha \approx \frac{4 N_v k_BT}{D_b} \int_{-\infty}^\infty \sinh^2\left[\frac{U(z)}{k_BT} \right]\,dz.
%\end{align} 

%Refinements to the CLS framework include extensions to regular solutions~\cite{imdrag:MendelevSrolovitz:2002} and gradient thermodynamics~\cite{imdrag:MaDregiaWang:2003} that become necessary at high solute concentrations, effect of co-segregation~\cite{imdrag:MendelevSrolovitzWE:2001}, interphase interfaces involving phase transformations~\cite{imdrag:HillertSundman:1976},  ceramic interfaces and more general interfacial complexions~\cite{gbseg:YanCannonBowen:1983,imdrag:GlaeserCannon:1986,gbm:PowersGlaeser:1998}, and two-dimensional models that factor in the in-plane fluxes~\cite{imdrag:RoyBauer:1975,imdrag:KorzhenevskiiBauschSchmitz:2006}. 

The CLS framework and its modern continuum and discrete refinements~\cite{imdrag:RoyBauer:1975,imdrag:HillertSundman:1976,imdrag:GlaeserCannon:1986,gbm:PowersGlaeser:1998,imdrag:MendelevSrolovitz:2002,imdrag:MaDregiaWang:2003,imdrag:KorzhenevskiiBauschSchmitz:2006,impdrag:Wicaksono:2013} highlight the main challenges in quantifying the solute-drag forces.
%and over the past few decades several meso- to continuum-scale studies have identified the thermodynamic and kinetic parameters necessary for quantifying the drag effect~\cite{imdrag:FanChen:1999, imdrag:MendelevSrolovitz:2002, imdrag:MaDregiaWang:2003}.   
Extracting the distribution $U(z)$ is challenging. In principle, the variation can be reconstructed from experimentally extracted segregation profiles at the interfaces~\cite{gbseg:Hondros:1965,gbseg:Meijering:1966,gbseg:Guttmann:1977,gbseg:Hondros:1977,gbseg:OguraVitek:1978,gbseg:Balluffi:1979,gbseg:Farber:2000,gbseg:Seidman:2002,imseg:HuberNeugebauer:2018}. 
%There has been considerable progress in high fidelity experimental characterization of interfacial segregation~\cite{gbseg:Seidman:2002, gbseg:Farber:2000}. 
However, 
%the shape of the solute-cloud is extremely sensitive to the form of the $U(\posVec, z)$, and 
these studies are limited in the atomic-scale resolution necessary to probe the dynamic nature of the atomic-scale interfacial structures at processing temperatures, and suffer from complications due to co-segregation effects. 
%More direct approaches that monitor in-situ interface motion are promising~\cite{gbm:Merkle:2002,gbm:Merkle:2004}, yet simultaneous monitoring of the interfacial structure, solute segregation profiles and interfacial motion remains a formidable challenge. 
%Atomic-scale computational frameworks %aimed at extracting interfacial segregation 
%can bypass some of these limitations~\cite{gbseg:Foiles:1989,gbe:NajafabadiSrolovitz:1993,gbseg:RittnerSeidman:1995}, yet they still rely on theoretical frameworks for interfacial segregation with attendant assumptions~\cite{gbseg:McLean:1957, gbseg:Lejcek:2010}.   
An independent quantification of intrinsic interface mobility and interface diffusivity is necessary (Eq.~\ref{eq:ExMobCLS}). Experimentally this is challenging since the level of sample purity and the solute excess at the interfaces required to estimate the drag forces are unknown, and 
%this is an issue given that 
there is growing evidence that even minute quantities of solutes (of the order of a few ppm) can modify the migration rates of grain boundaries~\cite{gbm:AustRutter:1959a,gbm:Upmanyu:1999,gbm:Gottstein:1998,book:HumphreysHatherley:1995,gbm:TrauttUpmanyu:2006}. 
 %In fact, the activation energies for grain boundary motion extracted via bicrystal experiments on high-purity metals are almost an order of magnitude higher than those extracted via all-atom computations of crystallographically similar yet completely pure bicrystals~\cite{}. The discrepancy is often attributed to the inordinately high driving forces employed in the simulations~\cite{gbm:Jhan:1990,gbm:Upmanyu:1998a,gbm:Schonfelder:1997,gbm:ZhangMendelev:2004,gbm:JanssensHolm:2006,imdrag:ZepedaRuizGilmer:2006}, yet 
Finally, both continuum and discrete frameworks breakdown at low driving forces (and therefore low velocities) where the transients driven by local solute-interface interactions become important~\cite{imdrag:LuckeDetert:1957,imdrag:MendelevSrolovitz:2002}. In this industrially relevant limit, the solutes are at near-equilibrium with the interface where dissipation via convective fluxes due to in-plane solute diffusion and shape fluctuations of the mechanico-chemically coupled interface become important.
%The interface is assumed to move at a constant velocity and under steady-state conditions and therefore ignore additional dissipation via convective . 
\begin{figure}[thbp] %  figure placement: here, top, bottom, or page
   \centering
   \includegraphics[width=\columnwidth]{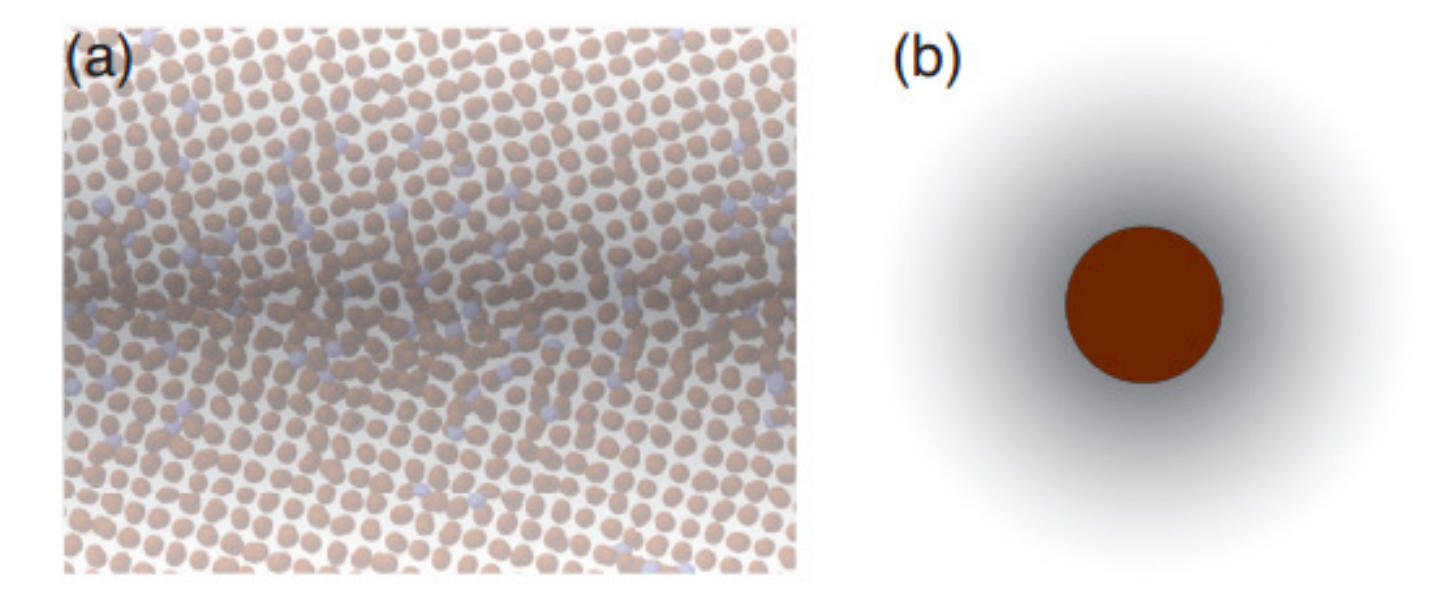} 
   \caption{(a) An atomic-scale configuration of a grain boundary in $\alpha$-Fe (red) with an equilibrium segregation of interstitial carbon (blue). The superposed grayscale contrast is a schematic illustration of the solute cloud whose motion is coupled to that of the boundary. (b) The equilibrium fluctuations of the grain boundary are analogous to diffusiophoresis of a Brownian particle in a binary fluid characterized by attractive interactions with the solvated solutes that form a cloud around the particle (shaded gray).}
   \label{fig:fig1}
\end{figure}

%Equilibrium fluctuations of pure interfaces have proven successful in extracting the mobilities~\cite{gbm:TrauttUpmanyuKarma:2006}.

Atomic-scale computational frameworks %aimed at extracting interfacial segregation 
based on equilibrium fluctuations of solute-loaded interfaces can bypass most of these limitations, enabling a direct quantification of the solute-drag effect. For example, the random walk method has been recently extended to quantify the effect of (a non-equilibrium distribution of) a varying number of solute atoms on the mobility of a grain boundary in a substitutional alloy~\cite{gbm:SunDeng:2014}. To make contact with experiments, a fundamental understanding of the effects of a near-equilibrium solute cloud is necessary, and to this end we first develop 
%a theoretical understanding.
%To this end, we
%Equilibrium fluctuations of solute-loaded interfaces capture all of these complications and can potentially enable the direct quantification of the solute drag effect. 
%first develop 
a Langevin framework for fluctuations of solute-loaded interface. We delegate a rigorous derivation of the shape fluctuations to a later study and present the limiting behavior for the short time-scale random-walk of the mean interface position for a simple homophase interface, i.e. a grain boundary. 
%(GB) that separates identical crystals which differ in their crystal orientation. 
%Past computational and experimental studies have shown the utility of random walk  of the mean boundary position as well as their shape fluctuations to extract their kinetic coefficient, the grain boundary mobility $M_i$~~\cite{gbm:TrauttUpmanyu:2006, gbm:SkinnerDullens:2010, gbm:FoilesHoyt:2006}.  
A  combination of numerical simulations are employed to validate the theory and the related spatio-temporal scalings that stem from the modified stochastic dynamics of the interface due to the presence of a bound yet diffusing solute cloud. We show the utility of this framework in quantifying the solute-drag force by performing all-atom computations of a grain boundary in the Fe-C system.

%and its dependence on thermodynamics and kinetics of the solutes and the interface.
%In this study, we explore the utility of equilibrium fluctuations of a solute-loaded interface for quantifying the solute-drag effect. 
%%%HEre
%Their atomic-scale computations show that, as in the pure case, the mean square displacement of the boundary increases linearly with time and the slope is a measure of the extrinsic mobility of the grain boundary. Here, we analyze this straightforward extension to an alloyed system by developing a Langevin framework for the random walk of a solute-loaded interface. 

%\section*{Results}

%The remainder of the article is organized as follows: we first revisit the theoretical underpinnings of the solute drag effect on crystalline interfaces, in particular the motion of interfaces with a segregating impurity cloud. Based on these frameworks, we propose a Langevin formulation for the dynamics of a fluctuating grain boundary, which naturally partitions the dynamics into qualitatively different and non-classical behavior across varying time-scales. Complementary numerical simulations with varying degrees of simplifications allow us to determine the validity of the theoretical predictions. We conclude with a discussion of the deployment of the framework for rational design of alloy systems, in particular stable nanocrystalline systems.

\section{Results}
\subsection{Langevin Formulation}
In order to capture the segregation-induced corrections to the fluctuations of the interface, it is necessary to account for the interplay between the thermal fluctuations of the interface and those of the solute cloud. Consider the interface profile defined as $h(\posVec, t) = \bar{h}(t) + \xi(\posVec, t)$, where $\bar{h}(t)$ is the spatially averaged boundary position and $\xi(\posVec, t)$ describes fluctuations in its shape. The coupled evolution of the interface and the solute cloud is modified by fluctuations at the interface and in the bulk,
\begin{subequations}
\begin{align}
\label{eq:GbDynamics}
\frac{h_t}{M_i} = \gamma \kappa - p_s  + \eta,
\end{align}
\begin{align}
\label{eq:concEvolution}
\frac{\partial c}{\partial t} =& \nabla\cdot \left[D_s\nabla c + \frac{D_s c}{k_BT} \nabla U  + \bForce \right].
%=& D \Delta c + \left[\frac{\partial D}{\partial z} +  \frac{D}{k_BT} \frac{\partial U}{\partial z} \right] \nabla c \nonumber\\
%&+ \frac{1}{k_BT} \left[ \frac{\partial D}{\partial z} \frac{\partial U}{\partial z} + D \frac{\partial^2 U}{\partial z^2} \right] c + \nabla \cdot \bForce.
%=& D \Delta c + A(z) \nabla c + B(z) c + \nabla \cdot \bForce. 
%\bar{h}_t \frac{\partial c(\PosVec,t)}{\partial z} + \nonumber\\
%&- \frac{D(\PosVec)}{k_BT} c(\PosVec,t) [1- c(\PosVec,t)] \nabla U(z) + \bForce(\PosVec, t) \Big]
\end{align}
\end{subequations}
 We ignore shear coupling between normal and tangential interfacial motion~\cite{gb:CahnTaylor:2004,gb:CahnMishin:2006,gb:KarmaTrauttMishin:2012,gbm:ZhangSrolovitz:2017}, for simplicity and also because solutes have a non-trivial effect in that they can uncouple the motion for boundaries that are otherwise coupled, and vice verse~\cite{gbm:WangUpmanyu:2014}. $\eta(\posVec, t)$ and $\bForce (\PosVec, t)$ are the interfacial and bulk noise terms that in the dilute limit are uncorrelated,
 %that, for dilute solutions, are still uncorrelated in space and time,
 % is still , i.e.
\begin{subequations}
\begin{align}
\langle\eta(\posVec, t)\eta(\posVec^\prime,t^\prime)\rangle = \frac{2k_BT}{M_i} \delta(\posVec-\posVec^\prime)(t-t^\prime),
\end{align}
\begin{align}
\label{eq:bulkNoise}
%&\langle\eta(\posVec, t)\eta(\posVec^\prime,t^\prime)\rangle = \frac{2k_BT}{\tilde{M}} \delta(\posVec-\posVec^\prime)(t-t^\prime),\nonumber\\
\langle q_i(\PosVec, t) q_j(\PosVec^\prime, t^\prime)\rangle &= 2D_s^B\frac{k_BT}{\partial\mu/\partial c} \delta(\PosVec-\PosVec^\prime)(t-t^\prime) \delta_{ij}\nonumber\\
&\approx2D_s^B c \; \delta(\PosVec-\PosVec^\prime)(t-t^\prime) \delta_{ij}.
\end{align}
\end{subequations}
The interfacial force due to relaxation is largely unmodified by the presence of the solutes with  $M_i\ne M_i(c)$ the intrinsic mobility of the pure interface with projected extent $L^{d-1}$ ($d$ being the dimension) that sets its effective diffusivity $D_i=M_i k_BT/L^{d-1}$. The bulk noise term captures the concentration fluctuations away from the interface that serves as the solute conservation condition~\cite{imdrag:Cahn:1962,sold:Karma:1993} that is based on the simplification $\partial\mu/\partial c=k_BT/c$.
%In the limit that the atomic-scale mechanisms associated with boundary motion remain unchanged, the intrinsic mobility does not depend on the presence of the solutes, $M_i\ne M_i(c)$. 
%where we have again used the dilute ideal solution , .  
The concentration fluctuations 
%have an indirect effect on the interface evolution effect in that they 
modify the solute flux,
\begin{equation}
\label{eq:Impflux}
{\bf J}=\frac{D_s^b}{k_BT} \nabla \mu + \bForce,\nonumber
%= D \frac{\partial c}{\partial z} + \frac{D c}{k_BT} \frac{\partial U}{\partial z} + v c = v c_\infty,
%\left[1- c \right]
\end{equation}
which results in fluctuations of the shape of the solute cloud (Eq.~\ref{eq:concEvolution}), therefore the drag force $p_s({\PosVec, t})$. 
The boundary condition $c(|z-h(\posVec, t)|>\lambda, t) = c_\infty$ together with solute flux balance across the interface consistent with solute conservation completes the formulation.

\subsection{1D Model}
%with appropriate boundary conditions Far away from the moving interface, $U(|z-h(\posVec, t)|>\lambda, t)= 0$ such that
%\begin{align}
%\label{eq:BC}
%c(|z-h(\posVec, t)|>\lambda, t) &= c_\infty.
%\end{align}
%This condition together with 
The effect of shape fluctuations $\xi(\posVec, t)$ and the resultant interface curvature on the spatially averaged boundary position $\bar{h}_t$ can be integrated out in the limit of fast in-plane solute diffusion. 
%In this limit, the interface mean square displacement (MSD) of the spatially averaged profile $\langle\bar{h}^2\rangle$ of a solute-loaded interface whose dynamics is governed by Eqs.~\ref{eq:GbDynamics} ands \ref{eq:concEvolution}.  
%A convective contribution solute cloud adjusts by both in- and out-of-plane diffusion~~\cite{imdrag:RoyBauer:1975, imdrag:KorzhenevskiiBauschSchmitz:2006}. The in-plane component leads to a convective contribution to the overall drag force that can drive morphological instabilities in the moving boundary [REF], but we expect these effects to be less pronounced during equilibrium fluctuations. 
%This becomes a reasonable approximation as  not directly dependent on the curvature term in Eq.~\ref{eq:GbDynamics} is  due to the in-plane periodic boundary conditions enforced in the computational studies~~\cite{gbm:TrauttUpmanyuKarma:2006}.  
%In addition, the relative importance of the shape fluctuations can be controlled in the computations by reducing the size of the computational cell - the extent of interfacial fluctuations decreases with their in-plane area~~\cite{sold:Karma:1993}.
Then, the 1D stochastic evolution of the interface follows from spatially averaging Eq.~\ref{eq:GbDynamics} along the interface,
\begin{align}
\label{eq:GbDynamics1D}
\frac{\bar{h}_t}{M_i} = \frac{1}{A} \iint_A h_t(\posVec, t) dx dy=  \left( - \bar{p}_s  + \bar{\eta} \right),
\end{align}
where the noise term $\bar{\eta}(t)$ is averaged over the interface area and still uncorrelated. The drag force can be expressed as an average over its spatio-temporal variation $\bar{p}(z,t)$ away from the interface (Eq.~\ref{eq:mcForce}),
\begin{align}
\label{eq:dragForce1D}
\bar{p}_s (t) &\approx  \frac{1}{\lambda}\int_{-\lambda/2}^{+\lambda/2} \Gamma(z,t) \,\, \frac{dU(z)}{dz} dz=\frac{1}{\lambda}\int_{-\lambda/2}^{+\lambda/2} \bar{p}(z,t) dz.
%\bar{p}_s (t) &= \frac{1}{A\lambda} \int_{\Omega_A} \left(\int_{-\lambda/2}^{+\lambda/2} \Gamma(\PosVec,t) \,\, \frac{dU(\posVec,z)}{dz} dz \right) d\posVec\nonumber\\
%& \approx  \frac{1}{\lambda}\int_{-\lambda/2}^{+\lambda/2} \Gamma(z,t) \,\, \frac{dU(z)}{dz} dz\,.
\end{align}
%Here, $\Gamma(z,t)$ and $U(z)$ are the solute excess and interaction strength averaged over their respective in-plane variation, with the dominant variation along the interface normal. 
The temporal dependence of the solute excess arises from asymmetry in the distribution of the solute excess about the boundary, 
%which simplifies to
%\begin{align}
%\label{eq:concEvolution1D}
%\frac{\partial c}{\partial t} =& \frac{\partial}{\partial z} \left[D_s \frac{\partial c}{\partial z} + \frac{D_sc}{k_BT} \frac{dU}{dz}  + \bForce \right],
%=& D \Delta c + \left[\frac{\partial D}{\partial z} +  \frac{D}{k_BT} \frac{\partial U}{\partial z} \right] \nabla c \nonumber\\
%&+ \frac{1}{k_BT} \left[ \frac{\partial D}{\partial z} \frac{\partial U}{\partial z} + D \frac{\partial^2 U}{\partial z^2} \right] c + \nabla \cdot \bForce.
%=& D \Delta c + A(z) \nabla c + B(z) c + \nabla \cdot \bForce. 
%\bar{h}_t \frac{\partial c(\PosVec,t)}{\partial z} + \nonumber\\
%&- \frac{D(\PosVec)}{k_BT} c(\PosVec,t) [1- c(\PosVec,t)] \nabla U(z) + \bForce(\PosVec, t) \Big]
%\end{align} 
with the MSD of the average interface position given by
\begin{align}
\label{eq:MSD1D}
\frac{\langle\bar{h}^2\rangle}{{M_i}^2}=& \int_0^tdt^\prime\int_0^tdt^{\prime\prime} \langle \left[\bar{p}_s (t^\prime)-\eta(t^\prime)\right] \left[\bar{p}_s (t^{\prime\prime})-\eta(t^{\prime\prime}) \right]\rangle.
%=& \frac{2{M}k_BT}{A^2} \int_0^tdt^\prime\int_0^tdt^{\prime\prime}\delta(\posVec-\posVec^\prime)(t^\prime-t^{\prime\prime})  d\posVec d\posVec^\prime\nonumber\\
%= \frac{2M_ik_BT}{A} \int_0^tdt^\prime\int_0^t \delta(t^\prime-t^{\prime\prime}) dt^{\prime\prime}.
%=& \frac{2\tilde{M}k_BT}{A} t.
\end{align}

Equations~\ref{eq:GbDynamics1D}-\ref{eq:MSD1D} describe the 1D Brownian motion of a particle in an external potential related to the drag force, $\partial V/\partial z=-\bar{p}_s (z,t)$,
%\begin{align}
%V(z, t) &= \frac{1}{\lambda} \int_{-\lambda/2}^{+\lambda/2} \Gamma(z,t)\, U(z) dz\,.\nonumber
%\int \bar{p}_i(z(t)) dz.\nonumber\\
%\end{align}
with a corresponding Fokker-Plank equation for the probability $\rho(z,t)$ of finding the boundary at position $z$ and time $t$, 
\begin{align}
\label{eq:FP}
%\frac{1}{M_i}\frac{\partial \rho(z, t)}{\partial t} = - \frac{\partial \left[ \bar{p}_s (z,t)\, \rho(z, t)\right]}{\partial z} + k_BT \frac{\partial^2 \rho(z, t)}{\partial z^2}\,.\nonumber
\frac{1}{M_i}\frac{\partial \rho(z, t)}{\partial t} = - \frac{\partial}{\partial z} \left[ \frac{\partial V(z, t)}{\partial z}\, \rho(z, t)\right] + k_BT \frac{\partial^2 \rho(z, t)}{\partial z^2}\,.
\end{align}
%with $\partial V/\partial z=\bar{p}_s (z,t)$. 
A compact analytical solution is difficult as the external potential itself evolves diffusively in time. The framework does, however, allow us to explore the limiting behavior that yields insight into the nature of the fluctuations, and we mention some of the salient aspects before using numerical techniques for the solving the coupled equations. 

%\noindent
%{\it Adiabatic Limit}: 
%Under most processing conditions,
\subsection{Scaling analyses}
In most polycrystalline alloys, the solute diffusivity is lower than the effective diffusivity of the interfaces, $D_i$. The extreme limit $D_s\ll D_i$ %the behavior can be described within 
corresponds to an adiabatic approximation in which the interface fluctuates in an external potential 
%of the form
\begin{align}
V(z,t) \propto \int \,\bar{p}_s[z/\zeta(t)] dz,\nonumber 
\end{align}
where $\zeta(t)$ is a continuous interpolating function that varies slowly on the time-scale of the motion of the interface.
%due to the combined effect of interface and bulk forces. 
It represents an approximation of the coupled dynamics of solute cloud between solutions to the quasi-static equilibrium configurations and allows us to explore the long- and short-time behavior of the system.
%, presented below.

\noindent
\subsubsection{Short time behavior}  
\label{sec:ShortTimeBehavior}
At time-scales much smaller than the solute diffusion scale $\tau_D\sim a^2/D_s$, the interface fluctuates in an effectively non-deformable solute cloud. Here, $a$ is the typical diffusion jump distance of the order of a few lattice parameters. The behavior is analogous to Brownian motion in a time-independent trapping potential $V(z)$ based on a near-equilibrium solute excess $\Gamma(z, t) \approx \Gamma(z)$.  The underlying Fokker-Planck equation has a stationary solution obtained by setting $\partial\rho/\partial t\rightarrow0$,
\begin{align}
%\label{eq:stSoln}
\rho_{st}(z) =\rho_{st}^0\,\, \exp\left(\frac{1}{k_BT} \int_0^z p_s^{eq}(z^\prime )dz^\prime\right) = \rho_{st}^0\,\exp\left[\frac{V(z)}{k_BT} \right],\nonumber
\end{align}
where $\rho_{st}^0$
%=\int e^{V^{eq}(z)/k_BT} dz$ 
is the normalization constant such that $\int \rho_{st} dz=1$. For displacements of the order of the width of the segregation profile, the external potential can be approximated as a quadratic potential $V(z)=-\frac{1}{2}\chi z^2$ with the constant $\chi=\partial p_s(z)/\partial z$ a measure of the 
%ensemble averaged 
drag force per unit interface width or its stiffness, 
\begin{align}
\label{eq:gradDragForce}
\chi \approx \langle \nabla\Gamma \,\nabla U +  \Gamma\,\nabla^2 U \rangle.
% +  \Gamma^{eq} \,\frac{d^2U}{dz^2}\,.
\end{align}
The resultant dynamics is the analogue of an overdamped Ornstein-Uhlenbeck process in a harmonic (friction) potential~\cite{bd:UhlenbeckOrnstein:1930} with
\begin{align}
\rho(z,t) &= \sqrt{\frac{\chi}{2\pi k_BT\, f(t)}}\, \exp\left[ -\frac{V(z)}{k_BT \, f(t)} \right] \, \text{and}\nonumber\\
\label{eq:MSD-OHProcess}
\langle \bar{h}^2 \rangle &= \frac{D_i}{M_i\chi}\,f(t), \,{\rm with}\, f(t) = 1-\exp(-2\chi M_i t).
\end{align}
The time dependence is captured by the exponentially increasing function $f(t)$ with a time constant associated with interface relaxation, $\tau_i=1/(2M\chi)$.   
%Linearization of the exponential in Eq.~\ref{eq:MSD-OHProcess} yields $\langle\bar{h}^2\rangle=2Dt$, that is the short-time behavior is initially classical as the interface fluctuates freely within the segregation potential well due to interfacial forces. Thereafter, 
%The cumulative effect of drag forces leads to sub-diffusive motion with 
The MSD is then necessarily non-classical at short times with the leading order correction directly proportional to the solute drag stiffness, 
%At larger displacements, the solute drag force reaches a maximum and then starts to decrease as the boundary begins to detach from the cloud. The quadratic assumption breaks down. 
%In order to gain a qualitative understanding of this effect, we employ an approximate solution for short-time dynamics of a Brownian particle within a generalized external potential $V(z)$,
\begin{align}
\label{eq:MSD-Approx}
\langle \bar{h}^2 \rangle &= 2D_it - \frac{2 D_i^2 L^d}{k_BT} \chi t^2 + \mathcal{O}(t^3),
%&+  \frac{D^3}{3k_BT} V^{\prime\prime}(z) t^3 + ,
%&\approx \frac{2 M_i k_BT}{A} t - \frac{2 M^2_i k_BT}{A^2} \left[\; \langle \Gamma^{eq} (z) \frac{dU(z)}{dz} \rangle \right]\, t^2,
\end{align}   
%The relation is valid for boundary displacements smaller that its width beyond which the boundary begins to detach from the cloud. 
and the deviation scales with the ensemble averaged drag stiffness. Following Eq.~\ref{eq:dragForce1D}, the net drag force is $\bar{p}_s=\lambda \chi/2$.
%Independent measures of the equilibrium solute excess at the interface $\Gamma^{eq}$ can then be used to extract out the average force per solute $\langle dU(z)/dz\rangle$, and therefore the average solute interaction strength, $\langle U(z) \rangle$.  

\noindent
\subsubsection{Intermediate behavior}
For time-scales of the order of the solute diffusion $t\sim \tau_D$, the segregation cloud deforms and becomes mobile, and the interface evolution is increasingly influenced by bulk forces.
 %and the dynamics of the interface can be expressed in terms of the random walk of the enveloping cloud. 
This has parallels with the Brownian dynamics of a particle in a binary solvent under an external force field that arises due to attractive surface interactions with the solutes (Fig.~\ref{fig:fig1}). The forces between the rapidly fluctuating interface and the cloud scale with their relative displacement,
%of the interface position and the maximum in the segregation profile, 
$\Delta h=|\bar{h} - \bar{h}_s|$, and the concentration fluctuations in the bulk lead to changes in the solute excess, $\Gamma (z)$. The overall evolution of the interface is due to their combined effect, $p_s[\Delta h(t), \Gamma (z, t)]$.
\begin{figure}[htbp] %  figure placement: here, top, bottom, or page
\centering
\includegraphics[width=0.8\columnwidth]{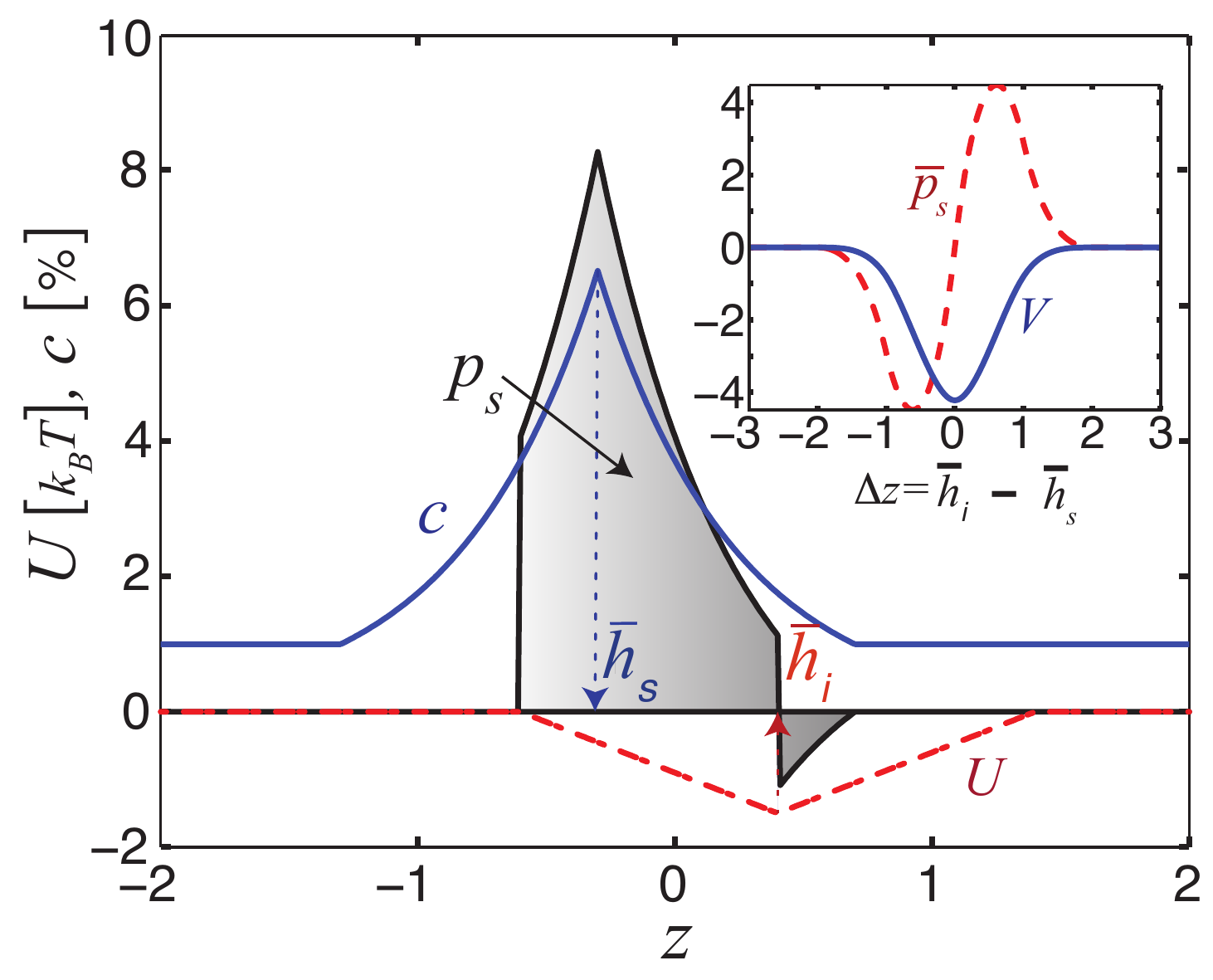}
\caption{Schematic showing the coupling between the dynamics of the interface and the solute cloud within a 1D Brownian dynamics (BD) simulation on a 1D lattice of unit cell length $a$, with solute-interface interaction strength
%(dimensionless) 
$U_0 = -1.5$\,eV, bulk solute concentration $c_\infty = 1\%$, and interface width $\lambda = 2a$. Here and elsewhere, the interface site density is fixed at $N_s=100/a^3$. The solute cloud is characterized by the peak in the equilibrium concentration profile (solid blue line), $\bar{h}_s = 0.3$ while the variation in the interaction strength is centered at the interface, $\bar{h}_i = -0.4$.  The shaded region represents the net drag force $\bar{p}_s$ that arises due to the relative displacement of the interface and the solute cloud, $\Delta z = \bar{h}_i - \bar{h}_s$. (inset) Plot of the drag force $\bar{p}_s$ and corresponding trap potential $V$ as a function of $\Delta z$. 
%The transients associated with changes in the shape of the solute cloud are ignored - see text for details. 
\label{fig:fig2}}
\end{figure}

Following the exponential sub-diffusion, the diffusion distance with respect to a static solute cloud ($\langle \bar{h_s}^2 \rangle=0$) saturates to a value that scales inversely with the drag stiffness, $\langle \Delta \bar{h}^2 \rangle \sim 1/\chi$ (Eq.~\ref{eq:MSD-OHProcess}). Thereafter,
%subsequent evolution of the solute cloud $\langle \bar{h}_s^2\rangle$ is then due to an additional fluctuating force that stems from the solute-interface interactions. In effect, the coupled system starts to diffuse due to . 
for small solute diffusivities $D_s/D_i\ll1$, the variance of the stochastic interfacial force scales as $\langle \bar{p}_s^2 \rangle \approx \chi^2 \langle \Delta \bar{h}^2\rangle\sim \chi$, where we have assumed that the fluctuations of the interface are rapid enough that they sample the entire solute cloud before it becomes mobile,
%\footnote{In situations where this approximation is invalid, the interfacial forces result an unbalanced force on the solute cloud over the times-scales it becomes mobile, resulting in a biased random walk of the coupled system.}
or $\langle \Delta h \rangle\approx 0$. 
%It also ignores higher order effects related to shape deformations of the cloud. 
The stochastic forcing of the solute cloud leads to a super-diffusive deviation from the classical response of an uncoupled solute cloud ($\sim D_st$) that scales as $(\bar{p}_s/D_i)^2t^2$ (Eq.~\ref{eq:FP}), resulting in a time-dependent effective diffusivity $\sim D_s + \chi t/D_i^2$. The sharpness of the initial recovery therefore increases with the drag force, and then eventually asymptotes to a response shaped by a steady-state balance between interface and bulk forces.  
%where we have ignored the much slower contributions due to the bulk diffusivity, $D_s^B$. 

\noindent
\subsubsection{Long time behavior}
At times $t\gg \tau_D$, the interface dynamics is continually modified by concentration fluctuations in the bulk as the interface absorbs and desorbs solutes. Fluctuations in the shape of the solute cloud represent a balance between this source-sink solute action and the interfacial forces. For low bulk diffusivities  $D_s/D_s^B\ll 1$ (e.g. substitutional solutes in crystalline alloys) and low interaction strength  $U\approx0$, the concentration fluctuations are essentially frozen and each solute acts as a local trap as the pure interface moves through their random dispersion. 
%We develop an intuitive understanding of the dynamics by first considering a system 
%If solutes do not segregate at the interface (), the pure interface fluctuates in a crystal with randomly dispersed solutes that serve as local traps. 
Diffusion in such generalized random media is anomalous and often sub-diffusive, characterized by a unique time exponent $\langle\bar{h}^2\rangle\propto t^\beta$ with $\beta<1$. Sinai diffusion represents an extreme limit characterized by particle motion in a quenched random external force field~~\cite{bd:Sinai:1982, bd:ComtetDean:1998}. At solute concentrations high enough to randomize solute-interface interactions, the effective drag force on the interface is spatio-temporally uncorrelated, $\langle \bar{p}_s(t^\prime) \bar{p}_s(t^{\prime\prime})\rangle = 2 U_0\delta\left[z(t^\prime)-z(t^{\prime\prime})\right]$. 
%\begin{align}
%\label{eq:SinaiCorr}
%\langle \bar{p}_s\left[z(t^\prime)\right] \bar{p}_s\left[z(t^{\prime\prime})\right]\rangle = 2 U_0\,\, \delta\left[z(t^\prime)-z(t^{\prime\prime})\right],\nonumber
%\end{align}
The interface dynamics is Sinai-like 
%provided the solute concentration is . 
%and the solutes do not accumulate at the interface. 
%Then, and the resultant Sinai diffusion leads to 
with logarithmic growth of the MSD, $\langle\bar{h}\rangle^2\propto \log^4(t)$,
%\begin{align}
%\langle\bar{h}\rangle^2\propto \log^4(t),\nonumber
%\end{align}
%with prefactors that scale with strength of the (non-accumulative) local solute-boundary interactions, $p_0$. 
that is sub-diffusive with an effective exponent that continuously decreases, or decelerating sub-diffusion. 
% due to the long-tailed distributions of the waiting times at the solute traps. 
On the other hand, in dilute solutions, 
%the interfacial dynamics occurs in a mostly pure matrix and 
solute trap events are rare and the net solutal force field acting on the interface is no longer uncorrelated. The interface motion is directed towards the solute traps interspersed with rapid diffusion between the traps. The distribution of wait times at the traps is set by the $z$-projected spacing between the randomly dispersed solutes and the strength of the traps. For heavy-tailed distribution of waiting times that can arise in dilute solutions, the dynamics corresponds to a (truncated) L\'{e}vy walker with super-diffusive dynamics, i.e. $\beta > 1$. 
\begin{figure}[htbp] %  figure placement: here, top, bottom, or page
\centering
\includegraphics[width=0.8\columnwidth]{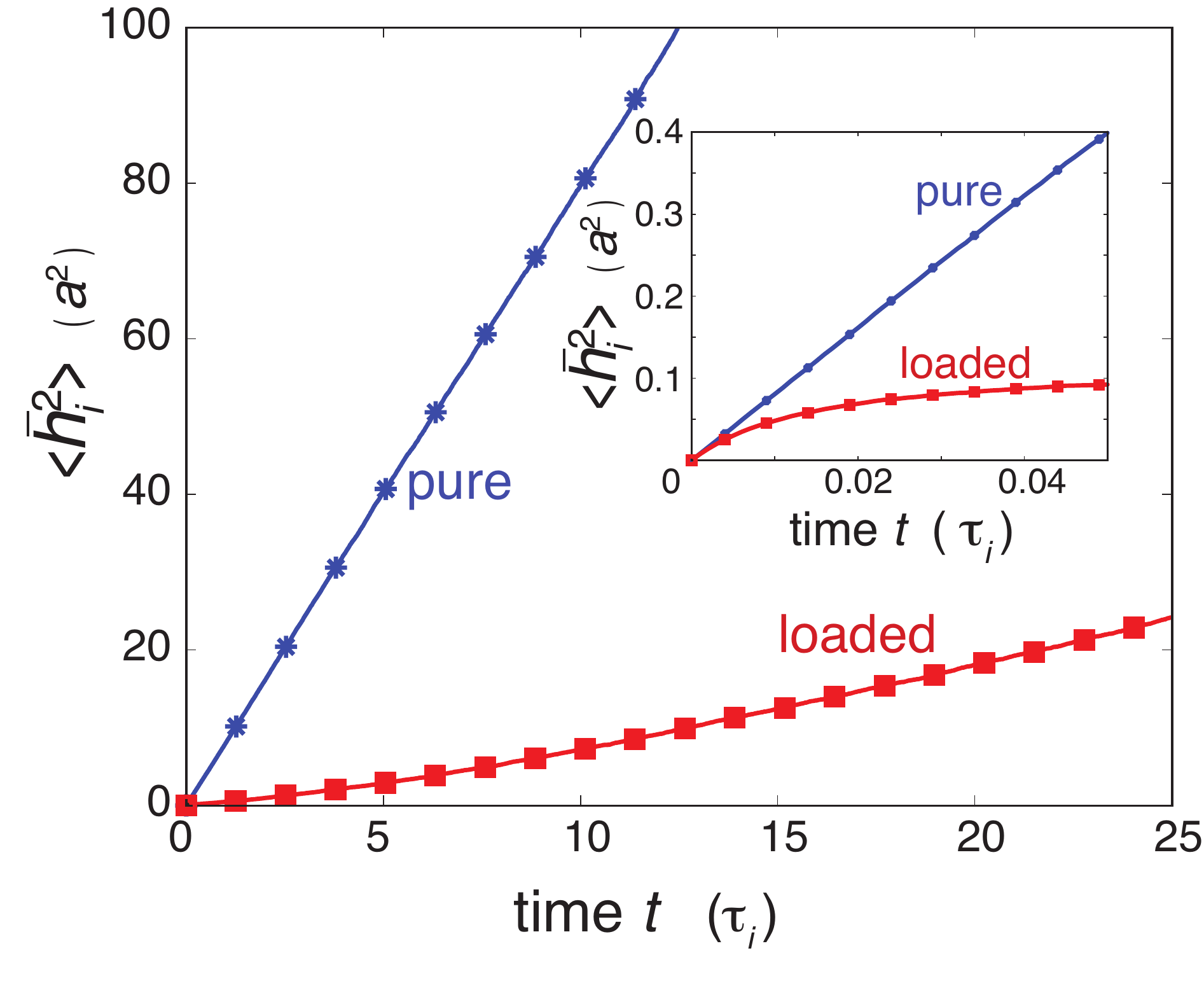}
\caption{Evolution of the MSDs of pure and solute-loaded interfaces in the BD simulations, with intrinsic interface mobility $M_i = 5\,a^2$/eV$\tau_i$ and $k_BT = 0.8$\,eV. For the solute-loaded interface, solute diffusivity $D_s = 0.01$\,$a^2/\tau_i$, $c_{\infty} = 1.0$ at\%, $U_0 = -1.5$\,eV and $\lambda = 2a$. The error bars are of the symbol size and not shown for clarity. (inset) A magnified view of the short time behavior.}
\label{fig:fig3}
\end{figure}

The dynamics of the solute-loaded interface lies in between these two limits. The segregation dilutes the effect of the local traps as it involves changes to an existing solute cloud due to the bulk forces. Additionally, a continuous exchange between solutes in the bulk and interface leads to a more normal distribution of the wait times, and we expect a classical behavior with an effective extrinsic diffusivity that self-consistently captures the balance between the interfacial and bulk forces. As an example, in the absence of bulk forces the spatially averaged interface velocity is proportional to the interfacial noise, $\bar{h}_t(t) = M_i \left( \bar{\eta} - \bar{p}_s \right) \approx M^\ast_i  \bar{\eta}$, and the MSD increases linearly with time,   
%$D^\ast_i=M^\ast_i k_BT/A$. 
%That is,
\begin{align}
\label{eq:longTimeSlope}
\langle \bar{h}^2 \rangle &= 2D_i^\ast t =  \frac{2 M^\ast_i k_BT}{L^d}\, t.
\end{align}
%&= \frac{1}{A} \iint_A h_t(\posVec, t) dx dy\nonumber\\
%&= M \left( \bar{\eta} - \frac{1}{A} \iint_A p_s dx dy \right) 
%&\bar{h} = M^{\ast}_i \int_0^t \bar{\eta}\,dt^\prime.
%\begin{equation}
%\bar{h} = M^{\ast}_i \int_0^t \bar{\eta}\,dt^\prime,
%\end{equation}
%Ignoring the evolution, the stationary case 
%Quite like CLS models of solute drag and its variants (Eq.~\ref{eq:ExMobCLS}), 
%the net effect of the drag force is to renormalize the mobility to its extrinsic value such that MSD increases linearly in time,
%In systems where the bulk solute diffusivity is negligible compared to that at the interface, the interface diffusion must occur over large distances to capture the bulk forces that now consist of solute source-sink action to and from the interface as it moves through the static solute dispersion. 
In substitutional alloys with low bulk diffusivities $D_s^B$, the extrinsic mobility entailes statistically significant interactions with bulk solutes for capturing the dependence of the bulk concentration $c_\infty$. In systems where the bulk diffusivity is comparable, we expect a quicker recovery of the classical random walk as it is facilitated by the bulk concentration fluctuations. In the extreme case where the bulk solute diffusivity is much larger (e.g. solid-fluid interfaces in liquid saline solutions~\cite{bd:AbecassisBocquet:2008}), the interface diffusivity is completely controlled by stochastic forces exerted by solute diffusion to and from the bulk. 
\begin{figure*}[htbp] %  figure placement: here, top, bottom, or page
\centering
\includegraphics[width=1.6\columnwidth]{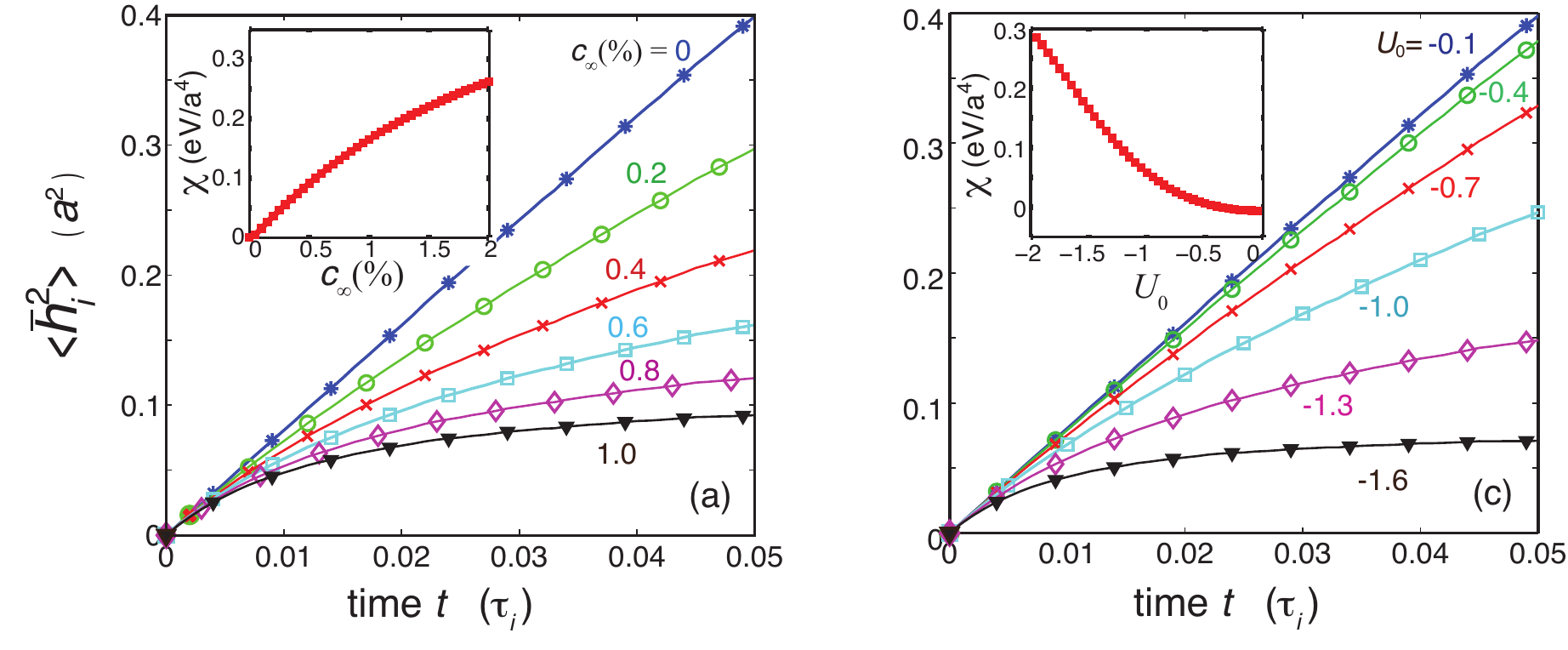}
\caption{Short-time evolution of the MSDs of interfaces in the BD simulations with varying (a) bulk solute concentration $0\le c_\infty \le 1.0\%$ and 
 %(b) solute diffusivity $D_s$ = 0.01, 0.2, 0.4, 0.6, 0.8 and 1; 
(b) solute-interface interaction strength $-0.1 \le U_0 \le -1.6$\,eV. The error bars are of the size of the symbols and not shown for clarity. 
%; (d) boundary width $\lambda$ = 0.8, 1.2, 1.6, 2.0 2.4 and 2.8. 
For both variations, the remaining variables are fixed at their reference values, $c_{\infty}$ = 1.0\%, $D_s= 0.01\,a^2/\tau_i$, and $U_0 = -1.5$\,eV. (insets) Plots of the drag force stiffness $\chi=\langle \nabla\Gamma\, \nabla U \rangle$ for each variation. The  average drag force $\bar{p}_s$ is the magnitude of $\chi$ for the prescribed interface width, $\lambda$ = 2a. 
\label{fig:fig4}
}
\end{figure*}

These scaling arguments reveal that the random walk of a solute-loaded interface is necessarily sub-diffusive at short times as the interface fluctuates in a frozen equilibrium solute cloud. Thereafter, the combination of slow solute diffusion and solute-interface interactions leads to a weakly super-diffusive intermediate recovery regime. At long times, 
%time-scales much larger than the solute diffusion time-scales, 
the interplay between bulk forces associated with concentration fluctuations and the interface forces results in a   gradual recovery of the classical linear evolution of the mean square interface position with a slope set by the extrinsic mobility of the interface.
% set by the interplay between interfacial and bulk forces.

\subsection{Numerical Analyses}
We use a series of computational frameworks with varying degrees of complexities to study the dynamics of a solute-loaded interface for varying energetic and kinetic parameters.  We first perform 1D Brownian dynamics simulations of two coupled random walkers. These simulations are performed in the quasi-static limit wherein transients in the shape of the solute cloud are ignored. These effects are captured in an on-lattice 2D Ising model computations with substitutional impurities. We validate the results via all-atom simulations of the short-time response of a grain boundary in $\alpha$-Fe in the presence of interstitial carbon. 

\noindent
\subsubsection{1D Brownian Dynamics of coupled random walkers}
%\begin{figure}[htbp] %  figure placement: here, top, bottom, or page
%\centering
%\includegraphics[width=\columnwidth]{Fig1a.eps} 
%\caption{The equilibrium concentration profile $c^{eq}(z)$ (top) and the boundary-%impurity interaction strength $U(z)$ (bottom) for the coupled Brownian dynamics %simulations.}
%\label{fig:fig1}
%\end{figure}
The overdamped Brownian Dynamics (BD) simulations are performed for two random walkers that represent the interface and the solute cloud with instantaneous coordinates $\bar{h}\equiv\bar{h}_i$ and $\bar{h}_s$, respectively (Fig.~\ref{fig:fig2}). For simplicity, we restrict the dynamics to 1D and ignore shape fluctuations of the interface and the solute cloud. 
The solute-interface interaction energy $U(z)$ is taken to be a triangular profile of width $\lambda$ centered at the interface (Methods and Eq.~S1). Its strength is scaled by its value at the interface center, $U_0$. 
%\begin{align}
%U[(z,t)/\bar{h}_i(t)]
%U[(z,t)] = & U[(z,t)/\bar{h}_i(t)] \\
%=  U_0\left(1-\frac{2|{z-\bar{h}_i}|}{\lambda}\right)
%\mathcal{H}\left(\frac{\lambda}{2}-|{z-\bar{h}_i}|\right)\nonumber
%\end{align}
 %$U_0$ is the thermodynamic variable that scales the strength of the interaction and $\mathcal{H}(x)$ is the Heaviside function that confines the interface-solute interactions within the interfacial width $\lambda$. 
At equilibrium, the two random-walkers are coincident, i.e. $\bar{h}_i=\bar{h}_s$ and the solute concentration profile (Eq.~\ref{eq:equibConc}) is symmetric about $\bar{h}_s$ and therefore $\bar{h}_i$.  
%\begin{align}
%c^{eq}(z, t) &= c_\infty \exp\left[-\frac{U[(z)/\bar{h}_s(t)]}{k_BT}\right]\nonumber
%\nonumber\\
%&= c_\infty \exp\left[-\frac{U[(z,t)/\bar{h}_i(t)]}{k_BT}\right].
%\end{align} 

The evolution of the two random walkers is implemented by numerically integrating the 1D Langevin equations (Eq.~\ref{eq:GbDynamics1D}) over a 1D lattice discretized with unit cell of length $a$  using a second-order stochastic Runge-Kutta (SRK) algorithm~\cite{bd:BrankaHeyes:1998} 
%\begin{align}
%\label{eq:BD_GB}
%\bar{h}_i[t+\Delta t] &= \bar{h}_i[t]  -  M_i \, \bar p_s [t + \Delta t/2] \Delta t + \theta_1,\nonumber\\
%\bar{h}_s{[t+\Delta t]} &= \bar{h}_s{[t]}+\dfrac{D_s}{k_B T} \,\bar p_s{[t+\Delta t/2]} \Delta t + \theta_2\,, 
%\end{align}
with uncorrelated white noises associated with the interface and bulk forces (Methods, Eqs.~S2-S3). The number density of solutes sites and temperature are both fixed at $N_s=100/a^3$ and $k_BT = 0.8$\,eV, respectively.     
%where $M$ is the intrinsic mobility of pure boundary and $D$ is impurity diffusivity. $\Delta t$, $k_B$ and $T$ are timestep, Boltzmann constant and temperature, respectively. 
For a net displacement between the interface and solute cloud $\Delta z=|\bar{h}_i - \bar{h}_s|$ induced by the fluctuations, the solute cloud is assumed to move rigidly with a concentration profile (Eq.~\ref{eq:equibConc}) based on an interaction strength centered at its instantaneous position $h_s(t)$. For simplicity, the solute diffusivity is assumed to be a constant throughout the 1D lattice. In effect, transients associated with the non-equilibrium solute profile are ignored, that is $U(z,t)\approx U[(z)/\bar{h}_s(t)]$ and $c(z, t)=c^{eq}(z)$. 
%\begin{align}
%U(z,t)\approx U[(z)/\bar{h}_s(t)]\nonumber, 
%c^{eq}(z, t)=c_\infty e^{-\frac{U[(z,t)/\bar{h}_s(t)]}{k_BT}}.
%= U_0(1-\frac{2|{z-\bar{h}_s(t)}|}{\lambda})\nonumber\\
%\mathcal{H}(\frac{\lambda}{2}-|{z-\bar{h}_s(t)}|).
%\end{align}
%$U(z,t)$ and $c(z,t)$ 
Using Eq.~\ref{eq:dragForce1D} we compute the instantaneous drag force
%\begin{equation}
%	\label{eq:BD_pi}
%	\bar p_s(t)= \int_{-\infty}^{+\infty} \Gamma(z,t)\,\frac{dU(z,t)}{dz}\, dz.\nonumber
%\end{equation}
%The definition 
%together with Eq.~\ref{eq:BD_GB}, 
%is used to numerically integrate the drag force 
that couples the two random walkers (Methods).
%\begin{equation}
%	\bar p_s {\left[t+\Delta t/2\right]}  = \frac{1}{2} (\bar p_s {[t]} +\bar p_s {[t+\Delta t]}), 
%\end{equation}
%with 
The results of the computations are analyzed by monitoring the interface MSD, $\langle \Delta \bar{h}_i^2 (t) \rangle=\langle [\bar{h}_i(t)-\bar{h}_i(0)]^2 \rangle$,   
%\begin{equation}
%	\label{eq:BD_MSD}
%	\langle \Delta \bar{h}_i^2 (t) \rangle=\langle [\bar{h}_i(t)-\bar{h}_i(0)]^2 \rangle,
%\end{equation}
where $\langle \cdots \rangle$ denotes the ensemble average over trajectories within independent simulations.  

%The equilibrium concentration profile within the solute cloud is based on 
%\begin{equation}
%	\label{eq:BD_c}
%	c(z,t)=c_{\infty} e^{-\frac{U(z,t)}{k_BT}}, 
%\end{equation}
%an interaction strength centered around the impurity position,
 %Here we assumed $\bar{h}_{i} $ is the 
Figure~\ref{fig:fig2} shows the configurations of the interface $\bar{h}_i = 0.3a$ and the solute cloud $\bar{h}_s = -0.4a$ within a simulation with intrinsic interface mobility $M_i=5$\,$a^4$/eV\,$\tau_i$ and interface area $A=1/a^2$. The reference values for the energetic and kinetic parameters are $c_\infty= 1.0\%$, $D_s=0.01$\,$a^2/\tau_B$, $U_0 = -1.5$\,eV and $\lambda=2a$. 
%The effect of is systematically studied for  at varying solute concentration $c_\infty$, solute diffusivity $D_s$, strength of interaction $U_0$ and segregation profile width $\lambda$, with reference values . 
The shaded areas represent restoring and repulsive contributions to $\bar{p}_s(t)$ around the interface center (Eq.~\ref{eq:dragForce1D}). The former is always dominant and results in a net drag force that varies non-monotonically with the relative displacement of the two random walkers $\Delta z$. It increases almost linearly for small $\Delta z$, exhibits a maximum at a distance of the order of the width $\lambda$, and then decays non-linearly to zero (inset, Fig.~\ref{fig:fig2}). The corresponding external potential $V$ is plotted in the inset. 

%At short times where the solute cloud is almost frozen, $\bar h_s \rightarrow 0$, the asymmetry is effectively the net boundary displacement $\langle \Delta z \rangle \rightarrow z$. For displacements of the order of the boundary width $z\le \lambda$, the drag force and the potential can be described in terms of the gradient of the drag force $\chi$ that is approximately constant, $V= - \frac{1}{2}\chi z^2$ and  $\bar{p}_s (z) =\chi z$. These scalings are useful for interpreting the short-time behavior and for extracting the time-averaged gradient of the drag force $\chi$ (Eq.~\ref{eq:gradDragForce}). The long-time behavior is used to extract the extrinsic mobility of GB,
%\begin{equation}
%	\label{eq:BD_ExM}
%	M^{ex} = \frac{1}{2 k_BT} \frac{d \langle \Delta \bar{h}^2 (t) \rangle}{d t}. 
%\end{equation}

\begin{figure*}[htbp] %  figure placement: here, top, bottom, or page
\centering
\includegraphics[width=1.9\columnwidth]{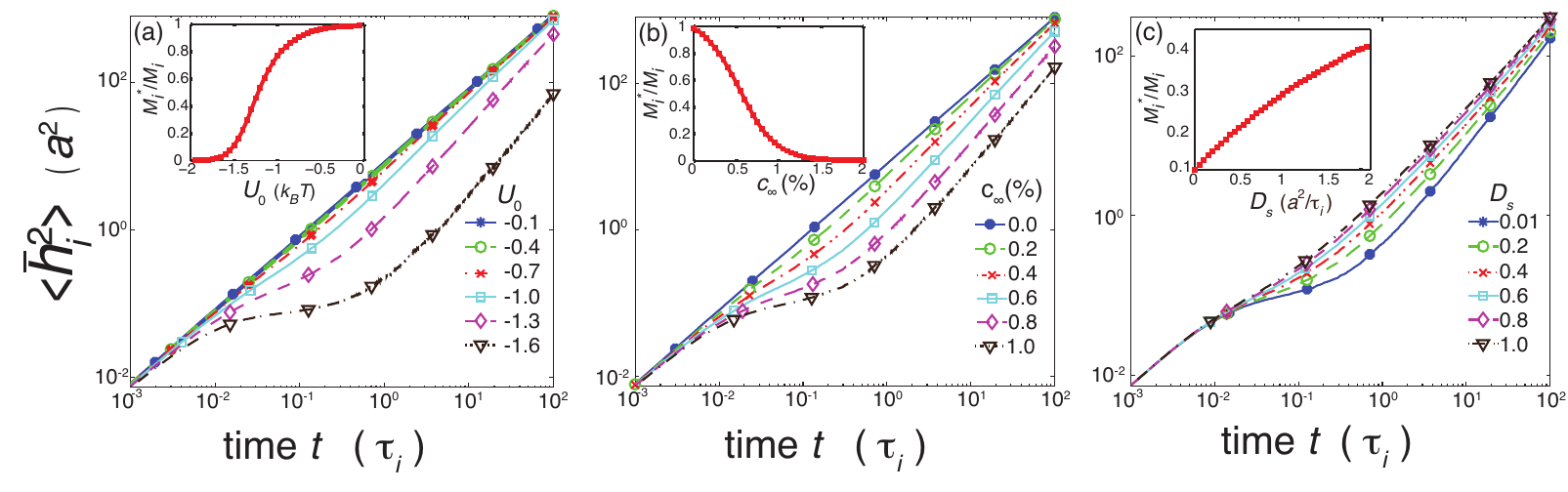}
\caption{Long time evolution of the interface MSD in the BD simulations with varying (a) $c_\infty$, (b) $U_0$ and (c)  $D_s$.
%  and (d) interface  width $\lambda$. 
The reference values for these variations are same as in Fig.~\ref{fig:fig4}. (insets) The effect of each variation on the ratio of the mobility of the solute-loaded interface to the pure interface, $M_i^\ast/M_i$ extracted from the slope $d\langle\bar{h}^2\rangle/dt$ in the range $60\,\tau_i<t<100\,\tau_i$. The error bars are of the size of the symbols.
%based on Eq. (\ref{eq:longTimeSlope}).
%as a function of (a)~$c_\infty$, (b)~$D_{\rm imp}$, (c)~$U_0$, (d)~$\lambda$ are shown on the inset. 
%$M^{\rm ex}$ results from the slope of MSD verse time at $t \in (60,100)$ based on Eq. (\ref{eq:BD_ExM}).}
%= 0(Pure GB), 0.2\%, 0.4\%, 0.6\%, 0.8\% and 1.0\%; (b) solute diffusivity $D_{\rm imp}$ = 0.01, 0.2, 0.4, 0.6, 0.8 and 1; (c) interaction strength of GB and impurity $U_0$ = -0.1, -0.4, -0.7, -1, -1.3 and -1.6; (d) grain boundary width $\lambda$ = 0.8, 1.2, 1.6, 2.0 2.4 and 2.8. The benchmark value for these series of comparison are $c_{\infty}^{\rm ref}$ = 0.01, $D^{\rm ref}$ = 0.01, $U_0^{\rm ref}$ = -1.5 and $\lambda^{\rm ref}$ = 2. The ratio of extrinsic GB mobility to intrinsic GB mobility  $M^{\rm ex}/M$ as a function of (a)~$c_\infty$, (b)~$D_{\rm imp}$, (c)~$U_0$, (d)~$\lambda$ are shown on the inset. $M^{\rm ex}$ results from the slop of MSD verse time at $t \in (60,100)$ based on Eq. (\ref{eq:BD_ExM}).}
\label{fig:fig5}}
\end{figure*}

%------Short timeFigure

%Short time behavior 
The temporal evolution of the MSDs of pure and solute-loaded interfaces are plotted in Fig.~\ref{fig:fig3}, for the reference values of $U_0$, $D_s$ and $\lambda$. The interface MSD increases linearly with time with a diffusivity consistent with prescribed interface mobility, $2D_i=2M_ik_BT/A=8\,a^2/\tau_i$. The evolution of the solute-loaded interface is initially non-linear and then asymptotes to a linear increase at long times, qualitatively in agreement with the theoretical predictions. The long-time slope is $2D_i^\ast\approx1.1\,a^2/\tau_i$, and based on Eq.~\ref{eq:longTimeSlope}, we get an order of magnitude decrease in the extrinsic mobility, $M_i^\ast=0.69$\,$a^4$/eV\,$\tau_i$. 
%The value can be also directly compared with the predictions of the CLS model for solute drag . 
As a comparison, for the triangular profile for $U(z)$ and reference values of $D_s$, $N_s$ and $\lambda$, the boundary-interaction parameter is $\alpha=134.2$\,eV\,$\tau_B/a^4$\,at\% and the CLS model predicts a value of $M_i^\ast=0.64\,a^4$/eV\,$\tau_i$ (Eq.~\ref{eq:ExMobCLS}), in excellent agreement with the simulations. The small discrepancy is largely due to lack of transients associated with bulk concentration fluctuations within a non-deformable solute cloud. 

At short times $t\le0.01\,\tau_i$ and for diffusion distances of the order of the lattice parameter (inset, Fig.~\ref{fig:fig3}), the MSD is highly sub-diffusive, in agreement with the theoretical predictions. A single-parameter exponential fit based on Eq.~\ref{eq:MSD-OHProcess} yields the spatio-temporally averaged drag stiffness, $\chi \approx 6.0\pm0.1$\,eV$/a^4$. For the prescribed interface width $\lambda=2a$, the magnitude of $\chi$ equals the average drag force $\bar{p}$ (Eq.~\ref{eq:dragForce1D}). Using the equilibrium solute excess $\Gamma(z)$ and $U(z)$ distributed over this width together with the distribution of $\Delta z$ observed in the simulations, the average value of $\chi = \langle \nabla\Gamma \,\nabla U \rangle = 5.8\pm0.4$\,eV$/a^4$. The agreement between the theoretical value increases for smaller times, suggesting that the small difference arises due to an increasing mobile solute cloud at larger times, and the non-monotonic variation in $p_s(z)$ at large $\Delta z$.
%(inset, Fig.~\ref{fig:1DBDSchematic}).  
\begin{figure*}[htbp] %  figure placement: here, top, bottom, or page
   \centering
   \includegraphics[width=1.8\columnwidth]{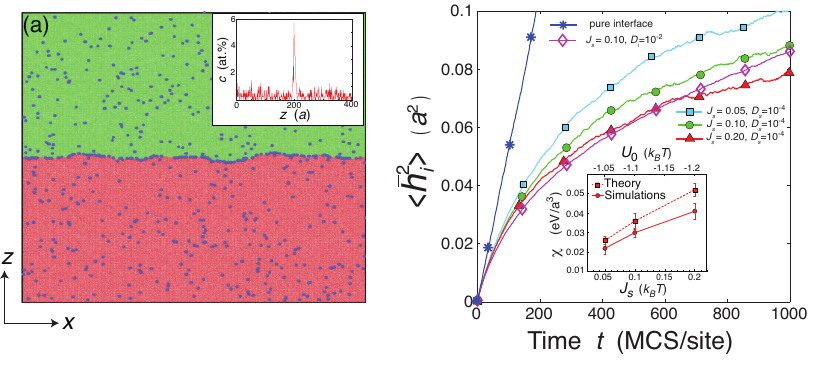} 
   \caption{(a) An equilibrated interface in an on-site square lattice employed for 2D kinetic Monte-Carlo (kMC) simulations. Red and green sites have spins (orientations)  $s_i=+1$ and $s_i= -1$, respectively. The solutes are colored blue with $s_i= 0$. (inset) The solute concentration distribution $c(z)$ along the interface normal for $c_\infty=0.3$\,at\%. 
 The solute-crystal atom bond energy is $J_s=1.0\,k_BT$. Here and elsewhere, the crystal and solute-solute bond strengths are fixed at $J_s=0.1$ and $J_{ss}=1.0$, respectively, in units of $k_BT=0.7$\,eV. (b) Short time evolution of the interface MSD for fixed $c_\infty=0.1$\,at\%, and for varying $J_s$ and solute diffusivities $D_s$. The error bars are of the size of the symbols and not shown for clarity.  The straight line corresponds to the pure interface. (inset) The effect of $U_0=-(J_s+J_c)$ on $\chi=\langle \nabla\Gamma\, \nabla U \rangle$ for $D_s=10^{-4}\,a^2 / {\rm (MCS/site)}$, extracted from the exponential sub-diffusive regime in the simulations (squares), and calculated based on the solute excess in the simulations $\langle\Gamma\rangle$ and the force per solute $dU/dz=-(J_s+J_c)/a$ in the model (circles, see Methods).}
   \label{fig:fig6}
\end{figure*}

The effect of variation in the bulk concentration $c_\infty$ and solute interaction strength $U_0$ with respect to the reference values on the short-time behavior is shown in Fig.~\ref{fig:fig4}a and~\ref{fig:fig4}b. 
In each case, the short-time behavior is exponentially sub-diffusive. As before, curve fits based on Eq.~\ref{eq:MSD-OHProcess} limited to $t\le0.05\,\tau_i$ yield the drag force gradient $\chi$,
%$=\langle \Gamma {dU}/{dz} \rangle$, 
and the results are plotted as insets in each corresponding panel.
%-----Unload vs Loaded with ref value 
%he cross-over was observed within the MSD curve using different grain width in  For large $\lambda$ like 2.0, 2.4 and 2.8, the slope of MSD increases as lambda increased. In contrast, for $\lambda$ smaller than 2.0, the slope decreases as $\lambda$ increased. This phenomenon was obviously reflect on the gradient of drag force as shown in the inset of Fig.~\ref{fig:fig3ST}d. The gradient of drag force reached a maximum value at approximate and then decreases along with increasing of $\lambda$.  %4 Lambda effect
Increasing $c_\infty$ leads to an increase in $\chi$  (Fig.~\ref{fig:fig4}a). Since $U_0$ and the interface width $\lambda$ are both fixed, the shape of the segregation profile as well as $dU/dz$ are unchanged such that the solute excess increases with $c_\infty$. We then expect the increase in $\chi (c_\infty)$ to be linear. 
%at the reference values. 
While this is indeed the case at small concentrations, the variation becomes weakly sub-linear with the concentration. The deviation arises due to the increasing confinement of the interface within a denser solute cloud. The resultant saturation of the MSD leads to corrections in the short-time behavior (Eqs.~\ref{eq:MSD-OHProcess} and \ref{eq:MSD-Approx}).
%Since the short-time cut-off for the curve fits is constant across all concentrations, the errors are expectedly larger for higher concentrations where the higher average drag force. 
Increasing the solute-interface interaction strength $U_0$ results in a dramatic increase in $\chi$ (Fig.~\ref{fig:fig4}b), as it enhances both the solute excess $\Gamma$ and the force per solute, $dU/dz$. The former is an exponential increasing quantity governed by the equilibrium solute segregation at the boundary (Eq.~\ref{eq:equibConc}), and is the dominant contribution. 
\begin{figure}[htbp] %  figure placement: here, top, bottom, or page
   \centering
   \includegraphics[width=0.8\columnwidth]{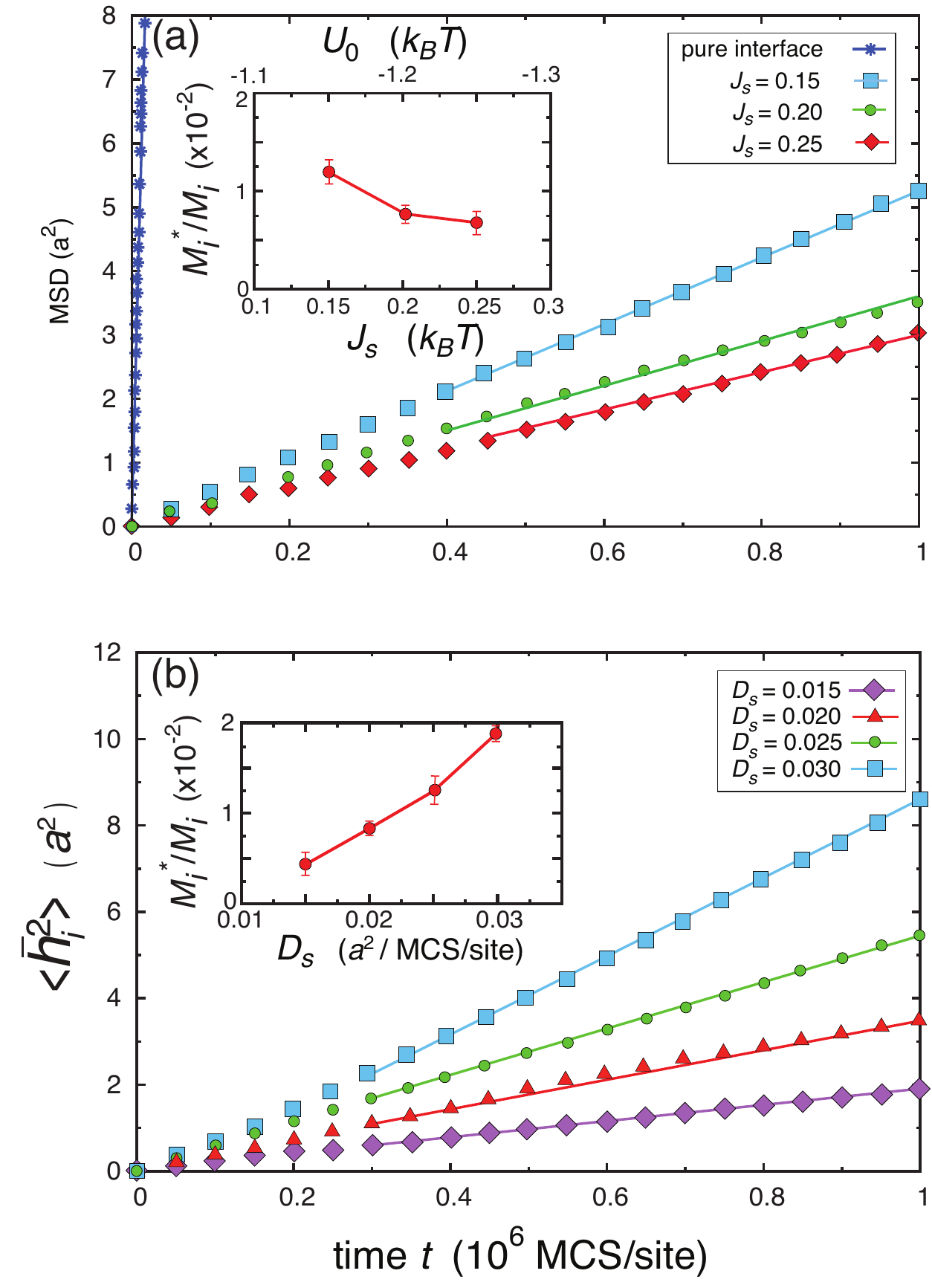} 
   \caption{Long-time evolution of the interface MSD in 2D kMC simulations for (a) varying $J_s$, at constant $D_s=0.01$\,a$^2$/ (MCS/site) and $c_\infty=0.1$\,at\%, and (b) varying $D_s$, at constant $J_s = 0.20$\, $k_BT$ and $c_\infty=0.1$\,at\%. Double logarithm versions of these plots are shown in Fig.~\ref{fig:SuppFig4}. (insets) The mobility ratio $M_i^\ast/M_i$ extracted from the slope $d\langle\bar{h}^2\rangle/dt$ as a function of (a) $J_s$ and (b) $D_s$.}
   \label{fig:fig7}
\end{figure}

The effect of solute diffusivity $D_s$ is expected to be minimal as the shape of the solute cloud is effectively rigid. 
%and the drag force and its gradient should be independent of the solute kinetics. 
%At small-times, . 
We do observe a weak dependence on $\chi$ ($\approx 20\%$) over a two orders of magnitude variation in $D_s$ (Fig.~\ref{fig:SuppFig1}a). The effect is similar to the $c_\infty$ variation in that a reduction in $D_s$ accelerates the saturation of $\langle \bar{h}_i^2\rangle$ and therefore leads to a deviations in the effective $\chi$. Decreasing the time-scale associated with the fit shows that the effect of $D_s$ is negligible. The interface width dependence $\chi(\lambda)$ is non-monotonic with a maximum at $\lambda \approx a$ (Fig.~\ref{fig:SuppFig1}b). For the triangular profile for $U(z)$, $\chi$ varies as $\exp(-U_0/\lambda)/\lambda$, and the net drag force $p_s\sim\lambda\chi$ increases exponentially. The small width behavior is dominated by exponential dependence of the solute excess while the decrease at large widths is due the decrease in gradient in the solute excess $d\Gamma/dz$ as well as the force exerted per solute, $dU/dz$. 

%Then, the net drag force averaged over all the solutes increases, which has the added effect of increasing the accumulation of errors in the curve fit. 

%The corresponding results with relationship of $c_\infty$ was almost exponential increased as shown on the inset of Fig.~\ref{fig:fig3ST}a.  %1 c_infinity effect
%Next, Figure.~\ref{fig:fig3ST}b suggests that as diffusivity of impurities $D$ was enhanced, at very short time beginning, the GB migration was almost consistent. After that, however, The GB coupled by impurities with lower diffusivity exhibits smaller mobility. Based on the MSD curve, we also extracted the gradient of drag force as described in the inset of Fig.~\ref{fig:fig3ST}b, which indicates that the greater impurities diffusivity results in smaller gradient of drag force. %2 Diffusivity effect
%In addition, we also compared the MSD curve among different value of . The deviation of MSD curves from the straight line like pure GB case was increased with the increasing $U_0$ value. As the interaction strength increased, the gradient of drag force increased exponentially as expected.  %3 U_0 effect 
%------Long timeFigure

Figure~\ref{fig:fig5} shows the evolution of the interface MSD over five decades of simulation time. The double-logarithmic plots for each parameter set exhibit a universal trend; following the initial exponential sub-diffusive response, we see a transition regime associated with a mobile solute cloud (the knee of each curve) wherein the time-exponent progressively increases and then settles into a weakly super-diffusive regime with $\beta>1$. Eventually, the behavior asymptotes back to classical diffusion with exponent $\beta\rightarrow1$. The long-time slope is used to extract the ratio of the extrinsic to intrinsic mobility ratio $M_i^\ast/M_i$. 
\begin{figure*}[htbp] %  figure placement: here, top, bottom, or page
   \centering
   \includegraphics[width=1.8\columnwidth]{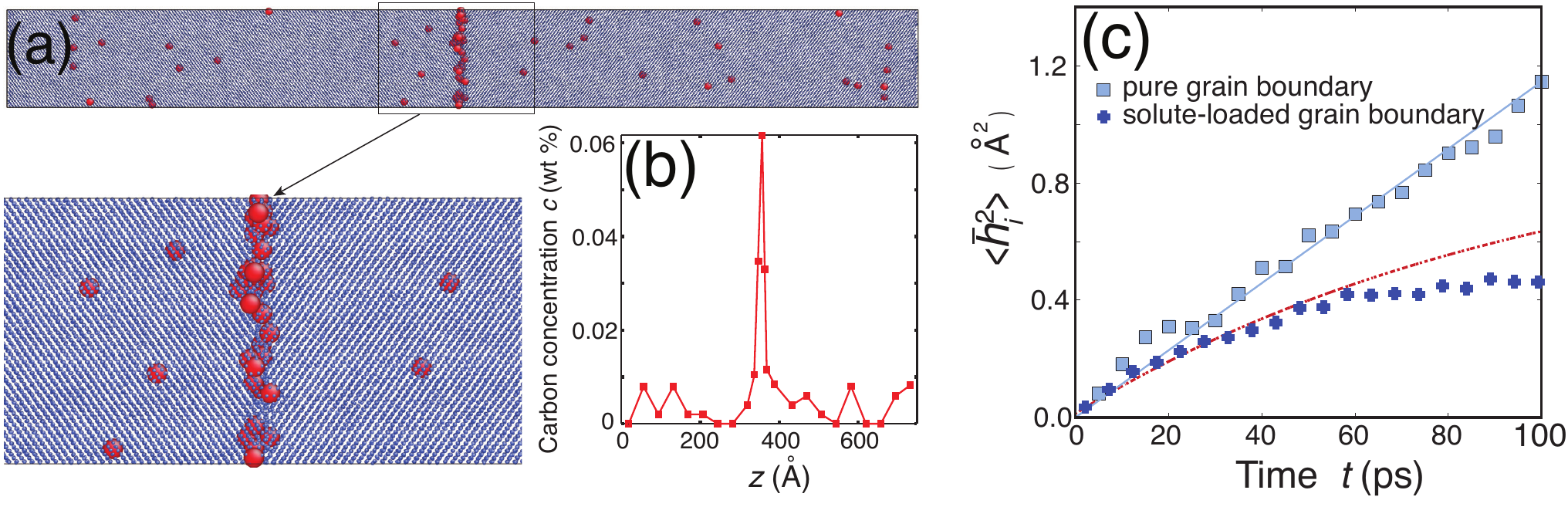} 
   \caption{ (a) Atomic configuration of one the equilibrated $\Sigma3\,(\bar{1}\bar{1}1)$ - $\theta = 109.5^{\circ}$ grain boundaries in $\alpha$-Fe. The bulk impurity concentration is $c_\infty=0.008$\,wt\%. Red and blue atoms denote carbon and iron, respectively.  The magnified view depicts the carbon distribution around the boundary in detail. The radius of carbon atoms is enhanced for visual clarity.  (b) The segregation profile $c(z)$ at $T=1000$\,K. (c) Short-time evolution of the grain boundary MSD with (circles) and without (squares) solutes  at $T=1000$\,K. The MSD is averaged over 100 independent simulations and the error bars are of the size of the symbols. The fits used to extract the mobility of the pure grain boundary and the drag force stiffness $\chi$ is also plotted. 
 \label{fig:fig8} 
 }
\end{figure*}

Increasing the thermodynamic parameters $U_0$ and $c_\infty$ enhances the sharpness of the recovery regime as well as the super-diffusive behavior thereafter (Fig.~\ref{fig:fig5}a-\ref{fig:fig5}b), consistent with the stochastic forcing of the solute cloud that leads to a time-dependent diffusivity that increases with the drag force $\chi$. The mobility ratio $M_i^\ast/M_i$ exhibits a sigmoidal decrease; as a measure of the solute-drag effect, we observe an order of magnitude decrease in the mobility for $U_0\approx -1.5$\,eV and $c_\infty\approx1.0$\,at\%. The interface width $\lambda$ has a similar similar effect on the mobility ratio (Fig.~\ref{fig:SuppFig2}), consistent with the exponential increase in the drag force (and therefore its gradient). The intermediate super-diffusive response decreases for extreme values of $\lambda$, mimicking the width dependence $\chi(\lambda)$ extracted from the short-time response, in accord with the scaling analysis. The solute diffusivity modifies the weakly super-linear increase in the intermediate regime (Fig.~\ref{fig:fig5}c), indicating that it is in a part a kinetic effect. Smaller diffusivities cause a separation of the scales between the rapidly fluctuating interface within a slowly evolving solute cloud, approaching a (truncated) L\'evy walker characterized by $\beta>1$. As such, the intermediate regime is increasingly super-diffusive for smaller diffusivities. At long times, the bulk forces drive the recovery of a classical random walk. The extracted mobility ratios increases with $D_s$, although the decrease is relatively modest and is consistent with the inverse dependence of the integral $\alpha$ in the CLS framework (Eq.~\ref{eq:ExMobCLS}).  On the other hand, $D_s$ has a negligible effect on the drag stiffness (Fig.~\ref{fig:SuppFig1}a), indicating that even though the solute cloud is non-deformable, the mobility reduction in this zero velocity limit remains in small part a kinetic effect. 

%Similarly, integrating both sides of Eq.~\ref{eq:drivingForceRel-RWExp} in space, we get
%\begin{align}
%\bar{h}_t &= M \iint_A \left[ \Gamma \left( h_{xx} + h_{yy} \right) + \eta - p_i \right] dx dy \\
%&= M \left( \bar{\eta} - \frac{1}{A} \iint_A p_i dx dy \right) = M \left( \bar{\eta} - \bar{p}_i \right).
%\end{align}
\noindent
\subsubsection{2D kinetic Monte-Carlo (kMC) simulations}
We next study the fluctuations of a solute-loaded interface in an on-site lattice using a 3-spin kMC simulations~~\cite{ising:Glauber:1963,imdrag:MendelevSrolovitz:2002}. Unlike the BD simulations, this model allows for self-consistent near-equilibrium deformations of the solute cloud. Figure~\ref{fig:fig6}a shows a bicrystal in a square lattice with a lattice parameter $a$. The crystal atoms have spins (orientations) $s_i=\pm 1$ while the substitutional solute atoms are prescribed spins $s_i=0$. The Hamiltonian of the crystal atoms is that of a ferromagnetic Ising model with bond energy $J_c=1$ (Methods and Eqs.~S4-S5). The bond strength between substitutional solutes and the crystal atoms $J_s>0$ results in a solute-interface interaction, shaped like a truncated triangle over an interfacial region of width $\lambda=3a$, with strength $U_0=-(J_s+J_c)$ (Fig.~\ref{fig:SuppFig3} and Table~1). The much rarer solute-solute interactions are prescribed a bond energy $J_{ss}=1$ that is sufficiently repulsive to prevent precipitation.       

Segregation at the interface is evident in the concentration profile $c(z)$ for $c_\infty=0.3$\,at\% and $k_BT=0.7$\,eV  (inset, Fig.~\ref{fig:fig6}a). Spin-flip dynamics with modifications for diffusing solutes reveal that the short-time fluctuations of the interface of the order of the lattice parameter are strongly sub-diffusive (Fig.~\ref{fig:fig6}b). The evolution of the pure interface is classical with a diffusivity $D_i=2.7\times10^{-4}\,a^2/\text{MCS\,site}^{-1}$. Single parameter fits to an exponential sub-diffusive behavior yield the drag force stiffness $\chi$ averaged over the interface width, and therefore the average drag force, $\bar{p}_s=3\chi/2$\,eV/a$^2$. For $c_\infty=0.1$\,at\% and solute diffusivity $D_s=10^{-4}\,a^2/\text{(MCS/site)}$ of the order of $D_i$, the interaction strength $U_0$ leads to an almost linear increase in $\chi$ (inset, Fig.~\ref{fig:fig6}b), similar to the trends in the BD simulations at higher values of $U_0$. Theoretical estimates for $\chi=\langle\nabla\Gamma\, \nabla U\rangle$ based on a constant  $dU/dz=-(J_s+J_c)/a$ for a flat interface and the gradient of the average solute excess at the interface $\langle \Gamma \rangle/3a$ extracted from the simulations are in good agreement and yield the same trend, although the simulation values are consistently smaller and the deviation increases with $U_0$. We attribute this deviation to the combination of co-segregation effects and roughness of the fluctuating interface that are naturally captured in the simulations. While small changes in the solute diffusivity have a negligible effect, for a solute diffusivity $D_s=10^{-2}\,a^2/\text{MCS\,site}^{-1}$ much larger than $D_i$, we see the drag stiffness increases from $0.3$\,eV/a$^3$ to $0.39$\,eV/a$^3$, and a rapid transition to the long-time behavior with a relatively steeper slope (Fig.~\ref{fig:fig6}b). The effect is in contrast to the minimal effect seen in BD simulations (Fig.~\ref{fig:SuppFig1}a), suggesting that the self-consistent deformations of the solute cloud modify the drag stiffness, and in turn the drag force.  

The long-time behavior for fixed $c_\infty=0.1$\,at\% and varying $U_0$ and $D_s$ is plotted in Fig.~\ref{fig:fig7}a and~\ref{fig:fig4}b. The solute diffusivities are chosen to be larger compared to the interface diffusivity in order to accelerate the response towards this regime. The behavior is classical in all cases, also evident in the double-logarithmic plots shown in Fig.~\ref{fig:SuppFig4}. The absence of the intermediate super-diffusive regime highlights the combined effect of slow interface diffusion and bulk forces on a compliant solute cloud. For solute-interface interaction strengths in the range $-1.3< U_0 < -1.1$\,eV, the drag effect is significant; we observe a two orders of magnitude reduction in the mobility ratio (Fig.~\ref{fig:fig7}a). Increasing the interaction strength and decreasing $D_s$ both lead to a non-linear decrease in the interface mobility (Fig.~\ref{fig:fig7}b). The trends are consistent with theoretical frameworks and BD simulations.

\subsubsection{Grain boundary in $\alpha$-Fe-C alloy}
Both BD and kMC simulations underscore the utility of the short-time interface fluctuations in quantifying the near-equilibrium solute drag stiffness $\chi$, and therefore the average drag force $\bar{p}_s$. We test the applicability of this approach for a grain boundary in a model interstitial alloy, a high symmetry $\Sigma3\,(\bar{1}\bar{1}1)$ - $\theta=109.47^{\circ}$ symmetric tilt grain boundary in the Fe-C system.   Carbon segregation to this grain boundary results in shear uncoupled motion at high temperatures $T>800$\,K~\cite{gbm:WangUpmanyu:2014}. The interatomic interactions are based on the Hepburn-Ackland empirical framework~\cite{intpot:HepburnAckland:2008} that accurately reproduces the thermodynamic and kinetic properties of interstitial carbon in $\alpha$-Fe (Methods). In order to capture the near equilibrium deformations of the solute cloud, we focus on the short-time response within a dilute alloy at a high temperature, that is $c_\infty= 0.008$\,wt\% and  $T=1000$\,K; the computationally intensive long-time behavior is delegated to a later study. 

The equilibrated atomic configuration of the segregated boundary of area $A=L_xL_y= 34.6$\,nm$^2$ is shown in Fig.~\ref{fig:fig8}a. Analysis of the carbon concentration profile (inset) yields the average boundary width $\lambda=2.2\pm0.02$\,nm beyond which the concentration approaches that in the bulk (Fig.~\ref{fig:fig8}b). This value is slightly larger than past studies on segregation of individual carbon atoms with grain boundaries in $\alpha$-Fe~\cite{seg:RhodesTschopp:2013}, likely due to co-segregation effects. In the absence of solutes, the MSD of the boundary in $\alpha$-Fe increases linearly with time~~\cite{gbm:TrauttUpmanyu:2006,fec:HoytUpmanyu:2010}, and the slope yields the diffusivity and mobility the grain boundary, $D_i= 5.1\pm0.08\times10^{-11}$\,m$^2$/s and $M_i=D_i A/k_BT=1.28\pm0.02\times10^{-7}$\,m$^4$/J~s respectively (Fig.~\ref{fig:fig8}c). Under these conditions, interface diffusivity is the almost twice the bulk diffusivity of interstitial carbon in this model system~\cite{intpot:HepburnAckland:2008,steels:JiangCarter:2003}, $D_s^B\approx3\times10^{-11}$\,m$^2$/s. On the other hand, the grain boundary diffusivity is at least an order of magnitude higher~\cite{steels:LuWen:2018}, $D_s^b\approx5\times10^{-9}$\,m$^2$/s.   

The short-time evolution of the solute-loaded grain boundary plotted in Fig.~\ref{fig:fig8}c is strongly sub-diffusive, similar to that in the BD and kMC simulations. The MSD saturates within $\approx60$\,ps and the average boundary displacement is less than the boundary width. While these are short scales, the boundary MSD is averaged over 50 independent equilibrium configurations of the interface and the solute cloud (Methods). Aided by the fast diffusion along the boundary, the boundary is able to rapidly explore near equilibrium configurations within these short spatio-temporal scales. The single parameter exponential fit is in excellent agreement with the response for For $t<60$\,ps, with a drag stiffness, $\chi=0.22\pm0.02\,{\rm eV/nm}^4$.
%=3.4\times10^{16}\pm0.11\,{\rm J/m}^4$. 
The average drag force over the boundary width is then $\bar{p}_s=\chi\lambda/2=38.5\pm1.7$\,MPa. 

\section{Discussion and Conclusions}
The theoretical analyses and the three distinct classes of computations clearly show that the random walk of a solute-loaded interface is non-classical at short times as it occurs within a slow moving solute cloud. The exponentially sub-diffusive fluctuations of the interface within the solute cloud yield the solute drag stiffness, and using independent measures of the interface width allows us to extract the average drag force (pressure). The extracted drag force is an average over near-equilibrium configurations of the solute cloud as well as self-consistent shape fluctuations of the interface. It therefore represents its (hitherto elusive) zero velocity limit, of direct relevance for the low driving forces typically during annealing phenomena. Our results show that for the $\Sigma3$ grain boundary in ferrite, for carbon concentrations as low as $c_\infty=0.008$\,wt\% the solute drag force is of the order of tens of MPa and is higher than typical driving forces in play during annealing of alloy microstructures. As a comparison, for average grain boundary energies $\gamma \sim 0.5$\,J/m$^2$ in metals and for micron-scale grain sizes, the curvature driving force $\bar{p}_s\sim2\gamma/R\approx1$. The effective motion of the boundaries at these grain sizes is therefore largely determined by the drag forces.

The ability to quantify the equilibrium drag force allows us to investigate the effect of thermodynamic and kinetic parameters such as bulk concentration and solute diffusivity, facilitating the development of comprehensive framework with predictive capabilities. The BD and kMC simulations point to some universal trends. The drag force is strongly influenced by the solute-interface interaction strength, and the effect of the solute diffusivity is indirect through near equilibrium deformations of the solute cloud. The diffusivity effect cascades to the interface mobility extracted at long times through an interplay between interfacial and bulk forces, and can be approximated as a force balance not unlike that at the heart of the CLS model (Eq.~\ref{eq:ExMobCLS}), wherein the effect of solute-interactions and its effect on interface shape fluctuations is absorbed within a now {\it velocity independent} drag force $\bar{p}_s\equiv \bar{p}_s(c_\infty, D_s, U_0)$ characterized by a slow moving solute cloud in equilibrium with the interface. This limit of small velocities and driving forces ($v/M_i\approx0$) leads to an extrinsic mobility that scales proportional to the interface velocity, $M_i^\ast=v/\bar{p}_s$. As an example, for the $\Sigma 3$ grain boundary in Fe-C, velocities of the order of $v\approx 1\,$mm/s lead to an extrinsic mobility $M_i^\ast\sim 10^{-12}$ orders of the order smaller than its intrinsic mobility.

The equilibrium drag stiffness and force serve as measure of the stability of nanocrystalline alloys in the system. Interfaces with average radius of curvature less than $R\sim2\gamma/\bar{p}_s$ can break free from the solute cloud. In the case of  $\Sigma3$ grain boundary in ferrite, $R\sim25$\,nm. The escape likely occurs at larger radii (and therefore grain sizes) as the high curvatures also reduce the solute excess~\cite{gbseg:Lejcek:2010}, which in turn increases the average grain boundary energy~\cite{gbseg:Kirchheim:2002}. The time-scales associated with the fluctuations with interfacial region are accessible for higher-fidelity computations using {\it ab-initio} dynamics, enabling rational search and design of interfaces and nanocrystalline alloy systems with improved thermal stability.

The extrinsic mobility can be directly extracted from the long-time regime. These are necessarily computationally intensive in order to capture the balance between interfacial and bulk forces; our ongoing efforts for the $\Sigma3$ grain boundary in $\alpha$-Fe at $T=1000$\,K show that the ensemble averages require times-scales in excess of 100\,ns. Longer computational times are required for substitutional alloys. Based on our BD simulations (Fig.~\ref{fig:fig5}c), the orders of magnitude slower solute diffusivities in these alloy systems imply that an intermediate superdiffusive regime mediates the transition to the long-time behavior. Much longer time-scales become necessary to capture the bulk forces that drive the transition to the long-time behavior. Extracting the drag forces from the short-time fluctuations offer a viable strategy around this time-scale limitation.

 An understanding of the hierarchical random walk of the interface within the coupled solute cloud has ramifications for stochastic dynamics in more general interacting systems, and therefore extends beyond the materials science context explored here. Although the solute dynamics is also stochastic, their preferential accumulation/depletion at the particle surface results in a media that adapts and evolves with the particle and is therefore strongly correlated. A similar situation arises when the fluctuating particles can sequester minority components from the medium, a classic example being ligand/antibody binding on moving vesicles and dynamics of biomacromolecules (e.g. DNA-binding proteins). Our study offers a framework for studying their dynamics in strong coupled, interacting media.

\section{Methods}
\subsection{Brownian dynamics simulations}
The Brownian dynamics simulations are performed on a 1D lattice with unit cell of length $a$. The simulations are performed for fixed interface mobility $M_i=5$\,$a^4$/eV\,$\tau_i$, interface area $A=1/a^2$, number density of solutes at the interface $N_s=100/a^3$, and thermal energy $k_BT=0.8$\,eV.
 
\subsubsection{Solute-interface interaction, $U(z)$}
The interaction between the solute and the interface is assumed taken to be a triangular profile of width $\lambda=2a$ centered at the interface
\begin{align}
\frac{U(z-\bar{h}_i)}{U_0}
%U[(z,t)] = & U[(z,t)/\bar{h}_i(t)] \\
=  \left(1-\frac{2|{z-\bar{h}_i}|}{\lambda}\right) 
\mathcal{H}\left(\frac{\lambda}{2}-|{z-\bar{h}_i}|\right) \tag{S1}
\end{align}
where $\mathcal{H}(x)$ is the Heaviside function that confines the interface-solute interactions to the interfacial width $\lambda$.

\noindent
\subsubsection{Dynamics}
The coupled evolution of the interface $\bar{h}_i$ and the solute cloud $\bar{h}_s$ is performed using a second-order stochastic Runge-Kutta (SRK) algorithm~~\cite{bd:BrankaHeyes:1998} 
\begin{align}
\label{eq:BD_GB}
\bar{h}_i[t+\Delta t] &= \bar{h}_i[t]  -  M_i \, \bar p_s [t + \Delta t/2] \Delta t + \theta_1, \tag{S2a}\\
\bar{h}_s{[t+\Delta t]} &= \bar{h}_s{[t]}+\dfrac{D_s}{k_B T} \,\bar p_s{[t+\Delta t/2]} \Delta t + \theta_2\,, \tag{S2b}
\end{align}
where $\Delta t$ is a time step much smaller than the relaxation time of the interface velocity $\tau_i\sim a^2/D_i$. For numerical stability and convergence, we choose $\Delta t=10^{-4}\,\tau_i$. The random numbers $\theta_1$ and $\theta_2$ are sampled from Gaussian distributions with zero mean value and variances 
\[
\langle \theta_1^2\rangle = 2M_i k_BT \Delta t, \quad{\rm and}\, \quad \langle \theta_2^2\rangle = 2D_s\Delta t,
\] 
respectively. The drag force that couples the two random walkers evolves as
%\begin{align}
%	\label{eq:BD_pi}
%	\bar p_s(t)= \int_{-\infty}^{+\infty} \Gamma(z,t)\,\frac{dU(z,t)}{dz}\, dz,\nonumber
%\end{align}
\begin{align}
\label{eq:BD_DragF}
%\bar{h}_i^\prime[t+\Delta t]_i &= \bar{h}_i[t]-M \bar{p}_s{[t]} \Delta t + \theta_1.\nonumber\\
\bar p_s {\left[t+\Delta t/2\right]}  &= \frac{1}{2} (\bar p_s {[t]} +\bar p_s {[t+\Delta t]})\,, \tag{S3}
\end{align}
where 
\begin{align}
\bar{p}_s{[t]} &\equiv\bar{p}_s(\bar{h}_i{[t]}, \bar{h}_s{[t]})\,\,\text{and}\,\, 
\bar{p}_s{[t+\Delta t]} \equiv \bar{p}_s(\bar{h}_i^\prime[t+\Delta t],\bar{h}_s{[t]}\nonumber
\end{align}
are evaluated based on the definition of $\bar{p}_s(t)$ (Eq.~7), and $\bar{h}_i^\prime$ is the trial displacement of the interface 
\begin{align}
\bar{h}_i^\prime[t+\Delta t]_i &= \bar{h}_i[t]-M \bar{p}_s{[t]} \Delta t + \theta_1.\nonumber
\end{align}

The simulations are typically performed for at least $10^6$ time steps to extract the long-time behavior. The MSD is averaged over $5\times10^4$ independent simulations and reported in units of $a^2$. For each variation ($c_\infty$, $U_0$, $D_s$, and $\lambda$), simulations are performed for 20 different values, with the remaining parameters fixed at their reference values (see main text).  

\subsection{2D Ising model with substitutional solutes}
{\it Energetics}: The Hamiltonian of the three spin bicrystal system ${s_i=\pm1, 0}$ is sum of the interaction energies between the crystal atoms, the solute-crystal atoms, and the solute-solute atoms, 
\begin{equation}
\label{eq:TotE}
E = E^{cc} + E^{si} + E^{ss}.\tag{S4}
\end{equation}
The crystal-crystal interactions for each of the $N_c$ crystal sites is based on the standard Ising model Hamiltonian, 
\begin{equation}
\label{eq:TotEss}
E^{cc} = - \frac{J_c}{2}  \sum_{i=1}^ {N_c}  \sum_{j=1}^{Z_c} s_i s_j,\tag{S5a}
\end{equation}
where $Z_c$ is the number of nearest neighboring crystal sites for the $i^{th}$ atom. The parameter $J_c$ is a positive exchange energy, related to the bond energy of each interacting crystal atom pair, and therefore sets the cohesive energy of the system as well as the interface energy. 

The interaction energy of each of the $N_{s}$ solute atoms with the crystal atoms $E^{si}$ is taken to be of the form~~\cite{imdrag:MendelevSrolovitz:2002}
\begin{equation}
\label{eq:TotEsi}
E^{si} = \frac{J_s}{2} \sum_{i=1}^{N_s}   | \sum_{j=1}^{Z_c} s_j |, \tag{S5b}
\end{equation}
%where $Z_c$ is the number of solvent atoms within the first nearest neighbor site of impurity atom $i$~~\cite{Mendelev2001}. 
where $J_s>0$ is the bond energy for each solute-crystal atom pair. 
The last term is Eq.~(\ref{eq:TotE}) is the impurity-impurity interaction energy $E^{ii}$, defined as
\begin{align}
\label{eq:TotEii}
E^{ss} = \frac{J_{ss}}{2} \sum_{i=1}^ {N_s}  \sum_{j=1}^{Z_{ss}}, \tag{S5c}
\end{align}
where $J_{ss}>0$ is the pairwise solute-solute bond energy and $Z_{ss}$ is the number of neighboring impurity atoms. 

\subsubsection{Solute-interface interaction strength} 
Figure~\ref{fig:SuppFig3} shows the bicrystal configuration with a single substitutional solute. The solute-interface interaction strength $U(z)$ can be expressed as
\begin{align}
U(z) = \left[E_s(z) - E_c(z)\right] - \left[E_s(z) - E_c(z)\right], \nonumber
\end{align}
where the term in the first square parenthesis is the energy change associated with adding a solute at a distance $z$ from the interface center, and the term in the second square parenthesis is the energy change associated with adding a solute in the bulk. 
% Requires the booktabs if the memoir class is not being used
\begin{table}[htp]
\caption{Parameters for calculation of solute-interface interaction energy for first neighbor interactions, $Z_c=Z_i=4$.}
\begin{center}
\begin{tabular}{|c|c|c|c|c|c|c}
\hline
$z$		& $E_s(z)$	& $E_c(z)$	& $E_s(\infty)$ 	& $E_c(\infty)$ 	& $U(z)$  \\
\hline
\hline
$a/2$	& $J_s$		& $-J_c$		& $2J_s$ 		& $-2J_c$ 	& $- (J_s + J_c)$ \\
\hline
$3a/2$	& $2J_s$		& $-2J_c$ 	& $2J_s$ 		& $-2J_c$ 	& $0$\\
\hline
\hline
\end{tabular}
\end{center}
\label{default}
\end{table}%

Table~1 shows the energy of a solute at distances $a/2$ and $3a/2$ from a flat interface (Fig.~\ref{fig:SuppFig3}a). The interactions are limited to the first nearest neighbors in the square lattice, i.e. $Z_c=Z_i=4$. Then, the $U(z)$ variation within the interfacial region of width $\lambda=3a$ is a truncated triangular profile,
\begin{align}
U (z) =
	\begin{cases} 
		-(J_s+J_c), & \text{if } z\le|a/2|\, \tag{S6}\\ 
		0 & \text{if } z \ge |3a/2| \,.\nonumber
	\end{cases}
\end{align}
%$U(a/2)=-(J_s+J_c)$ and $U(z\ge3a/2)=0$. 
Assuming a linear decay of $U(z)$ over the entire interfacial region, the average solute-interface force directed towards the interface is $dU/dz\approx -2/3 (J_s+J_c)/a$. In this study, $J_c=J_{ss}=1$, and therefore $dU/dz\approx -2/3 (J_s+1)/a$. 
%The cases with different value of attractive impurity-boundary interaction ($J_{\mathrm{si}}>0$) were investigated in this work. 

\subsubsection{Simulation geometry}
The simulations are performed within a rectangular simulation box of dimensions $L_x=L_z=400\,a$. Periodic boundary conditions are applied parallel to the interface ($x- $direction). In contrast, inverse periodic boundary conditions (i.e. $s_i \rightarrow -s_i$) are applied along the interface normal ($z-$direction), thereby eliminating the need for two interfaces and any artifacts due to interactions between the fluctuating interfaces. 

\subsubsection{Equilibrium segregation profile}
%------------------------3. modified GCMC algorithem -----------------------------------
Uncorrelated equilibrium solute distributions within the bicrystal are necessary for accurate ensemble averages of the interface fluctautions. They are generated using a semi-grand canonical Monte-Carlo simulations (SGCMC) adapted to on-site lattice dynamics~~\cite{surfseg:Foiles:1985}. These involve transmutation steps between crystal atoms and solute atoms, integrated within spin-flip dynamics (see below). The probability of accepting a transmutation trial is $\mathbf{min} (\Delta H, 1)$, and is based on the distribution function
\begin{align}
%\label{eq:GCMC_P}
	H=\frac{1}{N_c! N_s!} \exp \left( -\frac{E- \mu_c N_c - \mu_s N_s} {k_BT} \right),\nonumber
\end{align} 
where $\mu_c$ and $\mu_s$ are the chemical potentials of the crystal and solute atoms. Specifically, the probabilities $P_{s \rightarrow c}$ and $P_{c \rightarrow s}$ are 
\begin{align}
%\label{eq:GCMC-P2}
	P_{s \rightarrow i} &= \mathbf{min} \left[\frac{N_s}{N_c + 1}  \exp \left(-\frac{\Delta E + \Delta \mu}{k_BT} \right), 1 \right]\nonumber\\
	P_{c \rightarrow s} &= \mathbf{min} \left[  \frac{N_c} {N_s+1} \exp \left(-\frac{\Delta E - \Delta \mu} {k_BT} \right), 1 \right]\,,\nonumber
\end{align}    
where $\Delta E$ and $\Delta \mu=\mu_s - \mu_c$ are the change in the total system energy and relative chemical potential. 

The SGMC simulations are first carried out for single crystal to extract the relation between a fixed relative chemical potential $\Delta \mu$, consistent with the grand canonical ensemble, and the solute concentration in the vicinity of the desired $c_\infty$. SGMC simulations on a bicrystal with the same $\Delta \mu$ yield the segregation profile for a given $c_\infty$. 

\subsubsection{Spin-flip dynamics} 
The spin-flip evolution of the crystal atoms is performed using the standard Glauber dynamics of an Ising model~~\cite{ising:Onsager:1944, ising:Glauber:1963}. For a randomly selected site corresponding to a crystal atom, trial flips $s_i \rightarrow - s_i$ are accepted with a probability $\mathbf{min} \left[ \exp (-\Delta E /k_BT),1 \right]$.

\subsubsection{Solute diffusivity} 
The solute diffusivity is built on diffusion events wherein a solute exchanges its position with one of the (randomly selected) four nearest neighbor crystal atoms. For a randomly selected site corresponding to a solute atom, a diffusion event is accepted if $r \leq 4D_s$, where $r$ is a random number within the range $[0, 1]$. The event is accepted with a probability similar to the spin-flip event, $\mathbf{min} \left[\exp (-\Delta E /k_BT), 1 \right]$. As such, the diffusivity is a weakly dependent on the concentration - it decreases for a nearest neighboring solute. It increases/decreases for diffusion into/away from the interface region $\lambda\le3a$, with the net increase a balance between attractive forces between the solute and the interface and the higher solute concentration within the interface region. 

\subsubsection{Interface tracking and interface MSD}
The spatially averaged interface position in the bicrystal $\bar{h}(t)$ is related to the net spin in the system, normalized by the interface dimension. That is,
\begin{align}
\label{eq:GB location}
	\bar{h} (t) = \frac{N_c({s_i=1})-N_c({s_i=-1})}{2L_x},\nonumber
\end{align} 
where $N({s_i=1})$ and $N({s_i=-1})$ are the number of crystal atoms with spins up and down, respectively, and $L_x$ is the length of the interface. The temporal evolution of the interface MSD $\langle \bar{h}^2\rangle$ is averaged over 100 independent simulations and reported as a function of the Monte-Carlo steps per site, MCS$/$site.

\subsection{Atomic-scale simulations of grain boundary in $\alpha$-Fe}
\subsubsection{Interaction potential}
The Fe-Fe and Fe-C interactions are described using the empirical interatomic potential developed by Hepburn and Ackland~~\cite{intpot:HepburnAckland:2008}. The potential is well-suited for studying the energetics and kinetics of carbon interstitials in body-centered cubic iron. The study is limited to dilute carbon concentrations ($c_\infty=0.001$\,wt\%) and temperature $T=1000$\,K, where  precipitation of carbide particles (cementite) and the transformation to face-centered cubic (FCC) austenite are unimportant. 

\subsubsection{Simulation geometry} 
The simulation cell is periodic in-plane ($L_x=8.47$\,nm and $L_y=4.08$\,nm) and ends with free surfaces normal to the boundary ($L_z = 75.20$\, nm). Here we have $L_z$ almost 10 times larger than the other two dimensions to eliminate the interaction between free surface and grain boundary. To eliminate spurious effects due to segregation at the free surfaces, we employ $z$-edge regions with controlled carbon concentration set to $c_\infty$. The structure is locally relaxed at $0$\,K using conjugate gradient minimization and the temperature is gradually increased at zero pressure via isothermal-isobaric (NPT) molecular-dynamics (MD) simulations (No\'se-Hoover thermostat, time step of 2 fs). Atom deletion/addition interspersed with volume perturbations normal to the boundary plane (NPT MD) are employed to equilibrate the structure at $T=1000$\,K. The iterative procedure is implemented until energy converges to a minimum. 

\subsubsection{Equilibrium solute distribution}
The equilibrium carbon segregation profile is obtained using a grand-canonical Monte Carlo (GCMC) computations developed by the authors within the Large-scale Atomic/Molecular Massively Parallel Simulator, LAMMPS~~\cite{md:Plimpton:1995}. For computational efficiency, the Fe atoms are frozen initially during the GCMC simulations and interstitial carbon atoms are added, deleted and moved corresponding to the prescribed chemical potential. The segregation of carbon atoms leads to changes in pressure, and to that end addition/deletion of Fe atoms followed by NPT MD simulations are used to obtain the relaxed, energy minimum grain boundary structure. For more details on computational methodology, see Ref.~~\cite{gbm:WangUpmanyu:2014}.

\subsubsection{Grain boundary MSD} 
The molecular dynamics simulations were performed using LAMMPS. To get the ensemble averaged boundary MSD, we generate 100 independent configurations by performing canonical (NVT) Monte-Carlo simulations on the equilibrated segregation equilibrium profile. For computational efficiency, the trial moves are limited to the interfacial region.  Thereafter,  NPT MD simulations are again performed to relax any build of pressure, followed by NVT MD simulations. Each of the NVT simulations is run for at least 2\,ns. The boundary is tracked to a high degree of precision using a local orientation order parameter associated with the configuration of the Fe atoms~~\cite{gbm:TrauttUpmanyu:2006}.

\vspace{4 pt}
{\bf Acknowledgments}:   
This work is dedicated to the memories of Drs. John W. Cahn and Lasar S. Shvindlerman. It was supported by US Army Research Office (Award No. \#W3911 NF-14-1-0559). The computations were performed on {\it st}AMP and Discovery supercomputing resources at Northeastern University and Massachusetts Green High Performance Computing Center (MGHPCC). 

%\section*{References}

\begin{figure*} [htp]%  figure placement: here, top, bottom, or page
\centering
\includegraphics[width=1.8\columnwidth]{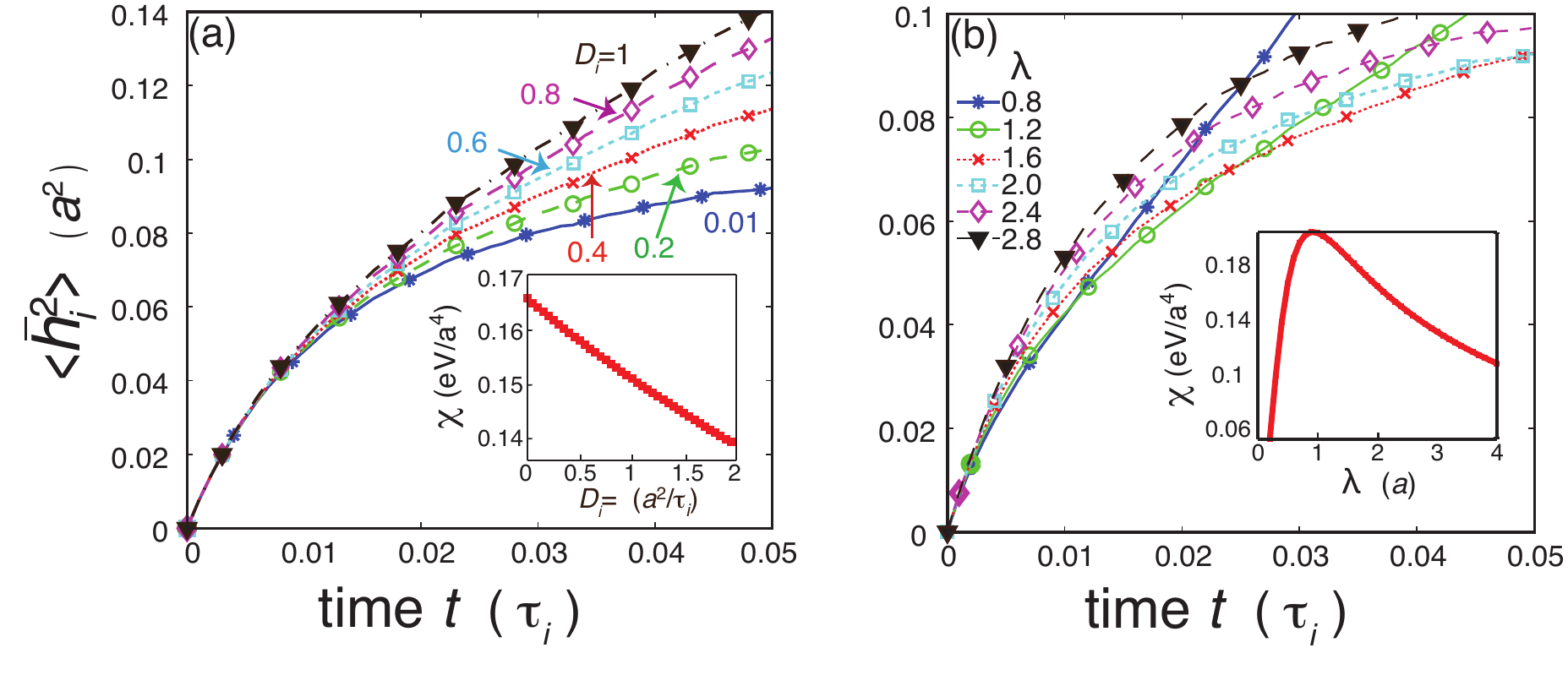}
\caption{Interface MSDs at short time-scales with varying (a) solute diffusivities  $0.01\le D_s\le 1$ and 
(b) interface widths $0.8\le \lambda \le 2.8$. For each variation, the remaining variables are fixed at the reference values. The insets show the plots of the drag force stiffness $\chi=\langle \nabla\Gamma \nabla U\rangle$ as a function of (a) $D_s$ and (b) $\lambda$, respectively.\label{fig:SuppFig1} }
\end{figure*}

\begin{figure*} %  figure placement: here, top, bottom, or page
\centering
\includegraphics[width=\columnwidth]{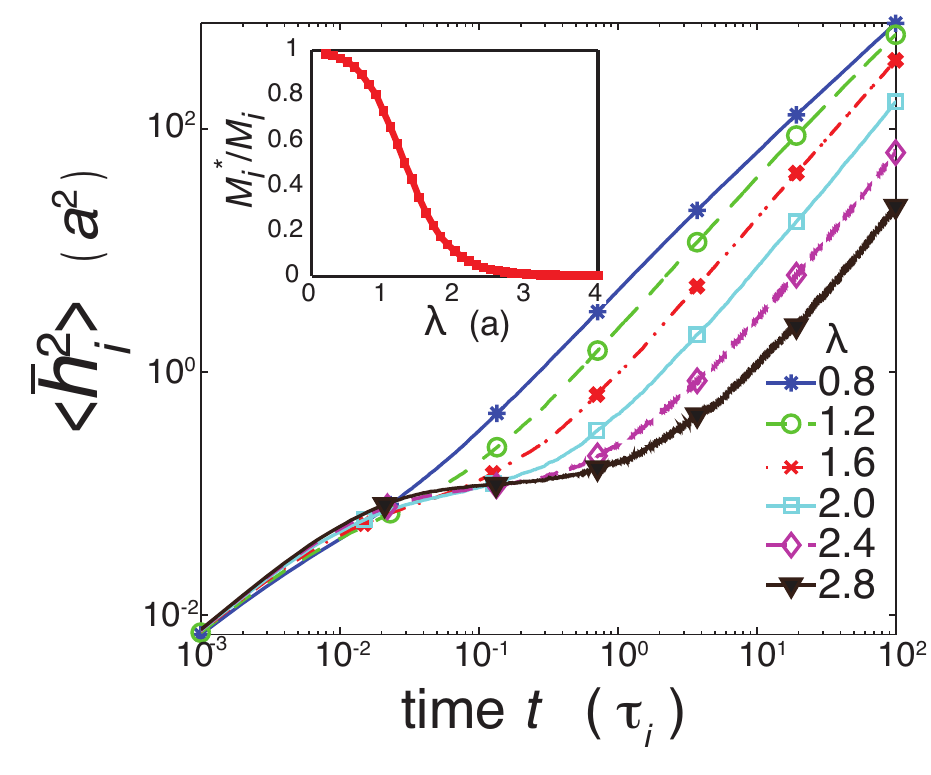}
\caption{Same as in Fig.~\ref{fig:fig5}, but for varying $\lambda$. The remaining variables are fixed at the reference values. (inset) The mobility ratio extracted in the range $60\,\tau_i\,<t<100\,\tau_i$.\label{fig:SuppFig2}}
\end{figure*}

\begin{figure*} %  figure placement: here, top, bottom, or page
\centering
\includegraphics[width=1.2\columnwidth]{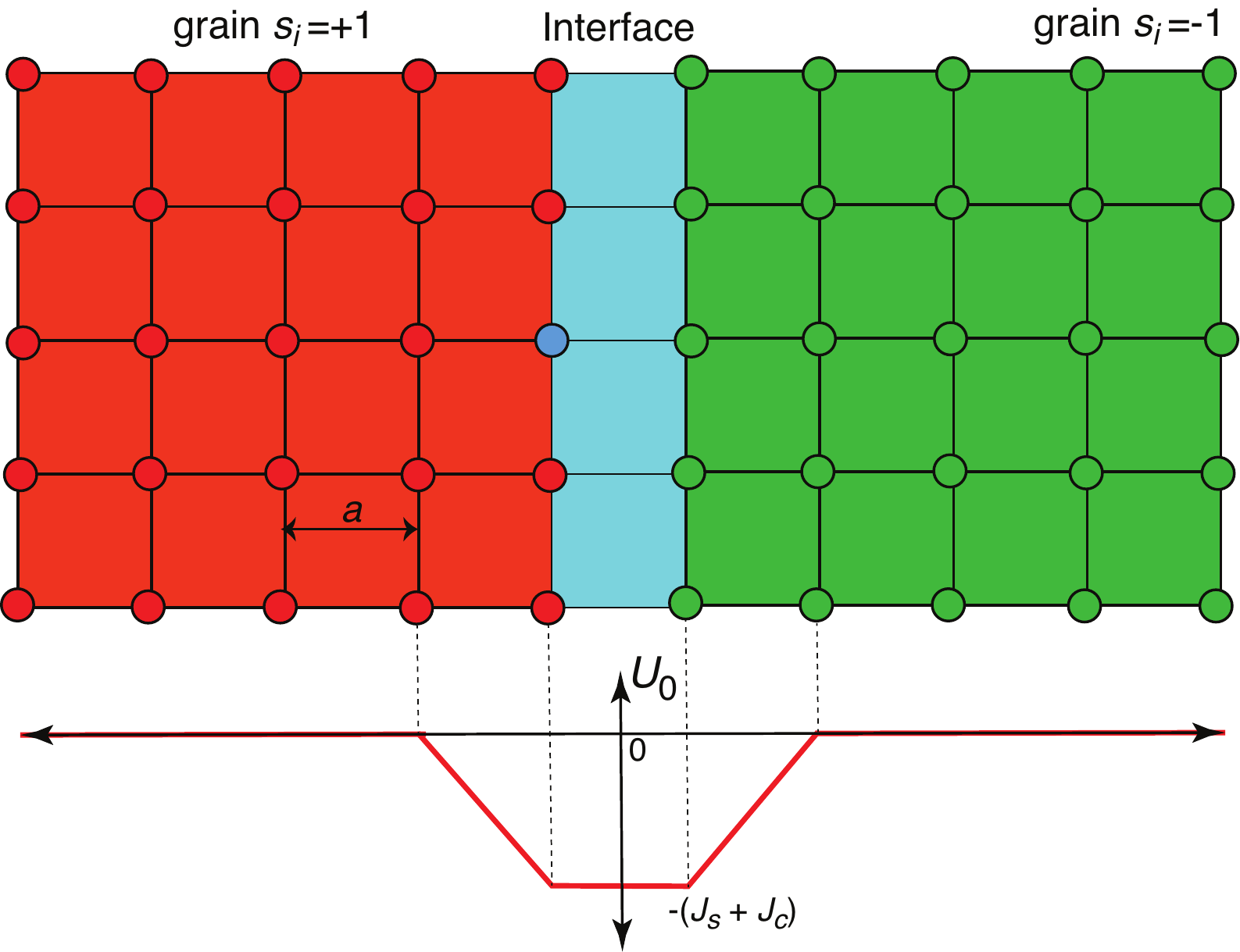}
\caption{Schematic illustrating the calculation of the solute-interface interaction strength for a flat interface in the 2D kMC model. The color scheme is the same as in Fig.~\ref{fig:fig6}. \label{fig:SuppFig3}}
\end{figure*}

\begin{figure*} %  figure placement: here, top, bottom, or page
\centering
\includegraphics[width=1.9\columnwidth]{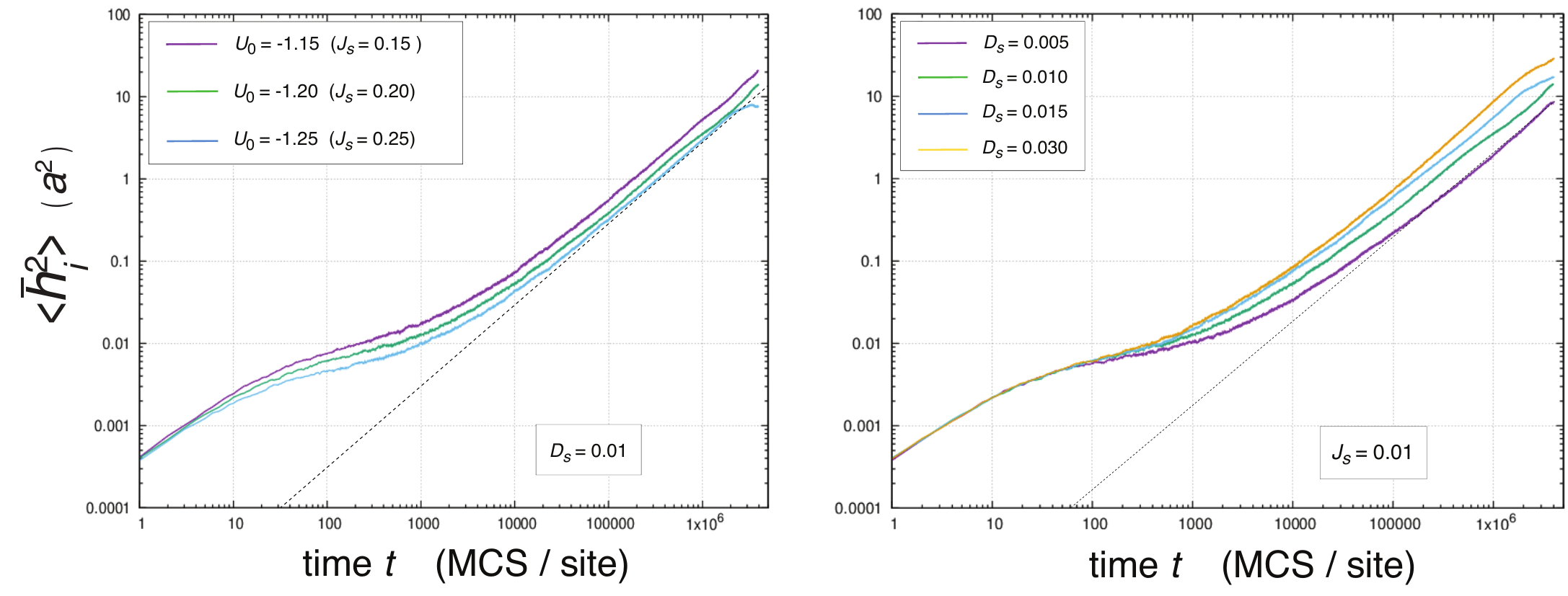}
\caption{Double logarithm plots of the long time evolution in the 2D kMC simulations shown in Fig.~\ref{fig:fig7}. The dotted line corresponds to the classical linear increase $\langle\bar{h}_i^2\rangle=2D_it$. \label{fig:SuppFig4}}
\end{figure*}

%\bibliography{references} 
\end{document}